\RequirePackage{amsmath}
\documentclass[runningheads]{llncs}
\usepackage[T1]{fontenc}
\usepackage{lineno}
\usepackage{graphicx}
\usepackage{amssymb}
\usepackage{amsmath}
\usepackage{graphicx}
\usepackage{breakcites}
\usepackage{subcaption}
\usepackage{placeins}
\usepackage{multirow}
\usepackage{booktabs}

\usepackage{todonotes}
\usepackage{xcolor}

\usepackage[hidelinks]{hyperref}

\DeclareMathOperator*{\argmaxB}{\arg\max} 

\begin{document}
\emergencystretch 3em
\title{Analyzing Modularity Maximization in Approximation, Heuristic, and Graph Neural Network Algorithms for Community Detection}
\titlerunning{Evaluating Modularity-Based Community Detection Heuristics}
\author{Samin Aref\inst{1}\orcidID{0000-0002-5870-9253}  \and \\
Mahdi Mostajabdaveh\inst{2}\orcidID{0000-0002-2816-909X}
}
\authorrunning{S. Aref et al.}
\institute{Department of Mechanical and Industrial Engineering, University of Toronto, M5S3G8, Canada \email{aref@mie.utoronto.ca}
\and
Department of Mathematical and Industrial Engineering, Polytechnique Montreal, H3T1J4, Canada
}
\maketitle              
\begin{abstract}
Community detection, which involves partitioning nodes within a network, has widespread applications across computational sciences. Modularity-based algorithms identify communities by attempting to maximize the modularity function across network node partitions. Our study assesses the performance of various modularity-based algorithms in obtaining optimal partitions. Our analysis utilizes 104 networks, including both real-world instances from diverse contexts and modular graphs from two families of synthetic benchmarks. We analyze ten inexact modularity-based algorithms against the exact integer programming baseline that globally optimizes modularity. Our comparative analysis includes eight heuristics, two variants of a graph neural network algorithm, and nine variations of the Bayan approximation algorithm. 

Our findings reveal that the average modularity-based heuristic yields optimal partitions in only 43.9\% of the 104 networks analyzed. Graph neural networks and approximate Bayan, on average, achieve optimality on 68.7\% and 82.3\% of the networks respectively. Additionally, our analysis of three partition similarity metrics exposes substantial dissimilarities between high-modularity sub-optimal partitions and any optimal partition of the networks. We observe that near-optimal partitions are often disproportionately dissimilar to any optimal partition. Taken together, our analysis points to a crucial limitation of the commonly used modularity-based methods: they rarely produce an optimal partition or a partition resembling an optimal partition even on networks with modular structures. If modularity is to be used for detecting communities, we recommend approximate optimization algorithms for a methodologically sound usage of modularity within its applicability limits. This article is an extended version of an ICCS 2023 conference paper (Aref et al., 2023) \cite{aref2023suboptimality}.


\end{abstract}
%
%
%


\section{Introduction}
Community detection (CD), the data-driven process of inductively partitioning nodes within a network \cite{schaub2017many}, is a core problem in computational sciences, particularly, in physics, computer science, biology, and computational social science \cite{fortunato2022newman}. Among common approaches for CD are the algorithms which are designed to maximize an objective function, modularity \cite{newman_modularity_2006}, across all possible ways that the nodes of the input network can be partitioned into communities. Modularity measures the fraction of edges within communities minus the expected fraction if the edges were distributed randomly; with the random distribution of the edges being a null model that preserves the node degrees. Despite their name and design philosophy, current modularity maximization algorithms, which are used by no less than tens of thousands of peer-reviewed studies \cite{Kosowski2020}, are not guaranteed to maximize modularity \cite{newman_equivalence_2016,kawamoto2019counting,meeks2020parameterised}. 

Modularity is among the first objective functions proposed for optimization-based community detection \cite{newman_modularity_2006,fortunato2016}.
Several limitations \cite{Guimer2004,fortunato2010community,fortunato2016,peixoto_2023} of modularity including the resolution limit \cite{fortunato_2007,schumm_bloom_2012,khomami_new_2018} have led researchers to develop alternative CD methods using stochastic block modeling \cite{Karrer_2011,sbm_2014,liu2021scalable,serrano2021community}, information theoretic approaches \cite{rosvall_2007,rosvall_2008}, and alternative objective functions \cite{surprise_2015,Aldecoa_2011_surprise,hamid_fast_2018,marchese2022detecting}. Modularity-based heuristics are the most commonly used methods for CD \cite{sobolevsky2014general,fortunato2022newman}. Besides modularity-based heuristics not guaranteeing the proximity to optimality, we do not know \cite{good_performance_2010,kawamoto2019counting} the extent to which they succeed in returning maximum-modularity (optimal) partitions or similar partitions. Recently developed alternatives to these heuristics are neural network-based algorithms \cite{sobolevsky2022gnn} and maximum modularity approximation algorithms \cite{aref2022bayan} which use different approaches for maximizing modularity. Unlike modularity-based heuristics, approximation algorithms provide guarantees on the proximity to optimality.  

Despite the availability of many modularity-based algorithms, the analysis of their performance in returning optimal partitions has not received sufficient attention \cite{good_performance_2010,kawamoto2019counting}. In a previous study, Good et al.\ used one algorithm and three metabolic networks to show that high-modularity sub-optimal partitions may substantially differ with each other \cite{good_performance_2010}. Our study is a continuation of a similar research path but it focuses on three fundamental and under-explored questions using an expanded scope of 104 networks, ten inexact algorithms, and an exact baseline method. 


The contributions of this study are therefore three-fold. (1) It investigates the extent to which modularity maximization algorithms return optimal partitions. (2) It quantifies the cost of sub-optimal partitions in terms of their dissimilarity to the closest optimal partition using three similarity metrics. (3) It evaluates the performance of ten modularity-based algorithms on a structurally diverse set of real and synthetic networks. Our comparative analysis reveals the differences between these algorithms. We investigate several methods from different perspectives and our results show that some methods have certain advantages over the others. This study aims to help practitioners select suitable modularity-based algorithm to use given the specifics of their use case. We do not claim or assume that maximum-modularity partitions represent best partitions; neither do we claim that modularity is a silver bullet for community detection. Throughout the paper, we use the terms network and graph interchangeably. After reviewing the ten algorithms and describing the methods and materials, we present the results in five subsections followed by a discussion of the methodological ramifications and future directions.

\section{Reviewing ten modularity-based algorithms}

We evaluate ten modularity-based algorithms known as Clauset-Newman-Moore (CNM) \cite{clauset_finding_2004}, Louvain \cite{blondel_fast_2008}, Reichardt Bornholdt with the configuration model as the null model (LN) \cite{rber_pots_2006,rb_pots_2008}, Combo \cite{sobolevsky2014general}, Belief \cite{zhang2014}, Paris \cite{paris_2018}, Leiden \cite{traag_louvain_2019}, EdMot-Louvain \cite{edmot_2019}, recurrent graph neural network (GNN) \cite{sobolevsky2022gnn}, and Bayan \cite{aref2022bayan}. We briefly describe how these ten algorithms use modularity to discover communities.

\textbf{CNM:}
The CNM algorithm initializes each node as a community by itself. It then follows a greedy scheme of merging two communities that contribute the maximum positive value to modularity \cite{clauset_finding_2004}.

\textbf{Louvain:}
The Louvain algorithm involves two sets of iterative steps: (1) locally moving nodes for increasing modularity and (2) aggregating the communities from the first step \cite{blondel_fast_2008}. Despite Louvain being the most commonly used modularity-based algorithm \cite{Kosowski2020}, it may sometimes lead to disconnected components in the same community \cite{traag_louvain_2019}.

\textbf{Leiden:}
The Leiden algorithm attempts to resolve a defect of the Louvain algorithm in returning badly connected communities. It is suggested to guarantee well-connected communities in which all subsets of all communities are locally optimally assigned \cite{traag_louvain_2019}.

\textbf{LN:}
The LN algorithm uses the same heuristic rules as the Leiden algorithm, but it supports weighted and directed graphs \cite{rb_pots_2008}.

\textbf{Combo:}
The Combo algorithm is a general optimization-based CD method which supports modularity maximization among other tasks. It involves two sets of iterative steps: (1) finding the best merger, split, or recombination of communities to maximize modularity and (2) performing a series of Kernighan-Lin bisections \cite{kernighan1970efficient} on the communities as long as they increase modularity \cite{sobolevsky2014general}.

\textbf{Belief:}
The Belief algorithm seeks the consensus of different high-modularity partitions through a message-passing algorithm \cite{zhang2014} motivated by the premise that maximizing modularity can lead to many poorly correlated competing partitions.

\textbf{Paris:}
The Paris algorithm is suggested to be a modularity maximization scheme with a sliding resolution \cite{paris_2018}; that is, an algorithm capable of capturing the multi-scale community structure of real networks without a resolution parameter. It generates a hierarchical community structure based on a simple distance between communities using a nearest-neighbour chain \cite{paris_2018}.

\textbf{EdMot:}
The EdMot-Louvain algorithm (EdMot for short) is developed to overcome the hypergraph fragmentation issue observed in previous motif-based CD methods \cite{edmot_2019}. It first creates the graph of higher-order motifs (small dense subgraph patterns) and then partitions it using the Louvain method to heuristically maximize modularity using higher-order motifs \cite{edmot_2019}.

\textbf{GNN:}
The GNN algorithm uses a recurrent graph neural network for maximizing modularity \cite{sobolevsky2022gnn}. It relies on a continuous optimization technique that considers current node's \textit{attachment}: continuous variable representing the cluster membership of a given node in a given community. In this algorithm, the attachments of nodes are combined with attachments of their neighbors. It starts with a random initial matrix of all attachments which is then updated iteratively to increase the modularity function using a recurrent graph neural network architecture \cite{sobolevsky2022gnn}. We have used two variations of the GNN algorithm: GNN-100 (suggested to be the fastest version) and GNN-25K (suggested to be a slow but very precise version) \cite{sobolevsky2022gnn}.

\textbf{Approximate Bayan:}
Unlike the algorithms discussed earlier, the approximate Bayan algorithm (Bayan for short) is an approximation algorithm for modularity maximization. Bayan is theoretically grounded by an IP formulation of the modularity maximization problem \cite{dinh_toward_2015}. For approximating an optimal solution to the IP problem, Bayan uses a branch-and-cut scheme \cite{aref2022bayan} while accounting for the gap between the upper bound and lower bound of the optimization problem. When the two bounds reach the desired approximation threshold (set by the user), the Bayan algorithm returns the partition with the highest modularity found alongside the maximum potential modularity gap (in percentage) that the returned partition may have from a globally maximum-modularity partition of the input graph. 

Except for Bayan and GNN, we use the Python implementations of the remaining eight algorithms (collectively referred to as heuristics) which are accessible in the Community Discovery library (\textit{CDlib}) version 0.2.6 \cite{rossetti2019cdlib}. For Bayan, we use \href{https://pypi.org/project/bayanpy/}{the \textit{bayanpy} version 0.7.6 library} in Python. And we use the GNN as implemented in its \href{https://github.com/Alexander-Belyi/GNNS}{public GitHub repository} referenced in \cite{sobolevsky2022gnn}. 

\section{Methods and Materials}

In this paper, we evaluate eight modularity-based heuristics \cite{clauset_finding_2004,blondel_fast_2008,rb_pots_2008,sobolevsky2014general,zhang2014,paris_2018,traag_louvain_2019,edmot_2019}, two variations of a graph neural network algorithm \cite{sobolevsky2022gnn}, and nine variations of the approximate Bayan algorithm \cite{aref2022bayan}. We quantify the extent to which these ten algorithms and their variations succeed in returning an optimal partition or a partition resembling an optimal partition. 


To achieve this objective, we quantify the proximity of their results to the globally optimal partition(s), which we obtain using an exact Integer Programming (IP) model \cite{brandes2007modularity,agarwal_modularity-maximizing_2008,dinh_toward_2015}. After describing the mathematical preliminaries, the IP model is discussed in Subsection \ref{ss:ip}.

\subsection{Modularity matrix of a graph}
Consider the simple (undirected and unweighted) graph $G=(V,E)$ with $|V|=n$ nodes, $|E|=m$ edges, and adjacency matrix entries $a_{ij}$. The modularity matrix of graph $G$ is represented by $\textbf{B}=[b_{ij}]$ whose entries are $b_{ij} = a_{ij}-{\gamma d_{i}d_{j}}/{2m}$. In this formula, $d_i$ represents the degree of node $i$ and $\gamma$ is the resolution parameter\footnote{Without loss of generality, we set $\gamma=1$ for all the analysis in this paper.}.

\subsection{Modularity of a partition}

For graph $G=(V,E)$, consider the partition $X=\{V_1,V_2, \dots, V_k \}$ of the node set $V$ into $k$ (non-overlapping) communities. The modularity function $Q_{(G,X)}$, proposed by Newman \cite{newman_modularity_2006}, maps each partition of a graph to a real number in the range $[-0.5, 1]$ according to Eq.\ \eqref{eq0}.

\begin{equation}
\label{eq0}
 Q_{(G,X)}= \frac{1}{2m} \sum \limits_{(i,j) \in V^2} \left( a_{ij} - \gamma\frac{d_id_j}{2m}\right) \delta(i,j)
\end{equation}

The modularity function $Q_{(G,X)}$ is based on the modularity matrix $\textbf{B}$ of graph $G$ and the partition $X$ applied on the node set of graph $G$. In Eq.\ \eqref{eq0}, the Kronecker delta, $\delta(i,j)$, is 1 if nodes $i$ and $j$ are in the same community according to partition $X$, otherwise it is 0.


\subsection{Modularity maximization}
The modularity maximization problem for the input graph $G=(V,E)$ involves finding a partition $X^*_{(G)}$ whose modularity is maximum over all possible partitions: $X^*_{(G)}=\argmaxB_{X}Q_{(G,X)}$. 

\subsection{Optimal and sub-optimal partitions}
For graph $G$, we refer to any partition that satisfies the definition of $X^*_{(G)}=\argmaxB_{X}Q_{(G,X)}$ as an \textit{optimal} partition (i.e., a maximum-modularity partition). Any partition that in not an optimal partition is a \textit{sub-optimal} partition. Different sub-optimal partitions $X_1,X_2$ of graph $G$ are distinguished based on their corresponding modularity values $Q_{(G,X_1)},Q_{(G,X_2)}$ as well as by their similarity to an optimal partition of $G$. If $G$ has multiple optimal partitions, we conservatively use the similarity to the optimal partition that is closest to the sub-optimal partition under evaluation. We use three different metrics for quantifying similarity to a reference partition which are described in \ref{ss:measures}.

\subsection{Sparse IP formulation of modularity maximization}
\label{ss:ip}
Consider the simple graph $G=(V,E)$ with modularity matrix entries $b_{ij}$, obtained using the resolution parameter $\gamma$. Consider the binary decision variable $x_{ij}$ for each pair of distinct nodes $(i,j),i<j$. The community membership of the nodes $i$ and $j$ is either the same (represented by $x_{ij}=0$) or different (represented by $x_{ij}=1$). Accordingly, Dinh and Thai \cite{dinh_toward_2015} have formulated the IP model for maximizing the modularity of input graph $G$ as in Eq.\ \eqref{eq1}. 

\begin{equation}
\label{eq1}
\begin{split}
  &\max_{x_{ij}} Q = \frac{1}{2m}   \left( \sum\limits_{(i,j) \in V^2 , i< j} 2b_{ij}(1- x_{ij}) + \sum\limits_{(i,i) \in V^2} b_{ii} \right) \\
&\text{s.t.}  \quad  x_{ik}+x_{jk} \geq x_{ij} \quad \forall (i,j) \in V^2 , i< j, k\in K(i,j) \\ 
& \quad \quad x_{ij} \in \{0,1\} \quad \forall (i,j) \in V^2 , i< j
\end{split}
\end{equation}

In Eq.\ \eqref{eq1}, $b_{ii}$ is the diagonal entry in row $i$ and column $i$ of the modularity matrix $\textbf{B}$ for graph $G$ which does not depend on the decision variables and is therefore separated from the off-diagonal entries. The optimal objective function value obtained from Eq.\ \eqref{eq1} equals the maximum modularity for the input graph $G$. An optimal community assignment is characterized by the values of the $x_{ij}$ variables in an optimal solution to the IP model in Eq.\ \eqref{eq1}. $K(i,j)$ indicates a minimum-cardinality \textit{separating set} \cite{dinh_toward_2015} for the nodes $i,j$.

Using $K(i,j)$ in the IP model of this problem leads to a more efficient formulation with $\mathcal{O}(n^2)$ constraints \cite{dinh_toward_2015} instead of $\mathcal{O}(n^3)$ constraints as in earlier IP formulations of the problem \cite{brandes2007modularity,agarwal_modularity-maximizing_2008}.
Solving this optimization problem is NP-hard \cite{brandes2007modularity,meeks2020parameterised}. To obtain the baseline of our comparative analysis, we use the \textit{Gurobi} solver (version 10.0) \cite{gurobi} to solve this NP-hard problem to global optimality for the small and mid-sized instances outlined in Subsection \ref{ss:data}. For each network instance, we first obtain the optimal partitions by solving the IP model in Eq.\ \eqref{eq1} using the Gurobi solver (version 10.0) and a termination criterion that ensures global optimality \cite{gurobi}. 

In the next step, we evaluate the ten modularity-based algorithms based on the proximity of their partitions to an optimal partition. On each network instance, we quantify the following for each algorithm (or algorithm variation): (1) the ratio of their output modularity to the maximum modularity for that network and (2) three measures of similarity between their output partition and an optimal partition of that network. These calculations lead to four values (GOP, AMI, RMI, and ECS) which are described in \ref{ss:measures}.

\subsection{Measures for evaluating the algorithms}
\label{ss:measures}

For a quantitative measure of proximity to global optimality, we define and use the \textit{Global Optimality Percentage} (GOP) as the fraction of the modularity returned by an algorithm for a network divided by the globally maximum modularity for that network (obtained by solving the IP model in Eq.\ \eqref{eq1}). 
In cases where an algorithm returns a partition with a negative modularity value, we set GOP=0 to facilitate easier interpretation of proximity to optimality based on non-negative GOP values. 

We use three measures to quantify the similarity of a partition to an optimal partition. Two of them (AMI and RMI) are grounded in information theory and are shown to be reliable measures of partition similarity \cite{jerdee2023normalized}. We use adjusted mutual information \cite{vinh_AMI} and normalize it symmetrically \cite{jerdee2023normalized}. The symmetrically normalized adjusted mutual information (AMI for short) \cite{vinh_AMI} is a measure of similarity between two partitions of the same network. We also use reduced mutual information \cite{newman2020RMI} and normalize it asymmetrically \cite{jerdee2023normalized}. The asymmetrically normalized reduced mutual information (RMI for short) \cite{newman2020RMI} is a measure of similarity between two partitions of the same network.

Unlike the commonly used \cite{roozbahani_community_2023,singh_disintegrating_2022,hamid_fast_2018,khomami_new_2018,sattari_cascade_2018,schumm_bloom_2012} yet problematic \cite{vinh_AMI,gates2019element,newman2020RMI,jerdee2023normalized} normalized mutual information (NMI) \cite{vinh_AMI}, AMI and RMI adjust the measurement based on the similarity that the two partitions may have by pure chance. AMI and RMI for a pair of identical partitions (or permutations of the same partition) equal 1. For two extremely dissimilar partitions, however, AMI and RMI take values close to 0.

To ensure the reliability of our results on similarities of partitions, we also use the Element-Centric Similarity (ECS) as a third measure of partition similarity \cite{gates2019element}. ECS differs from AMI and RMI in that it uses an alternative method for quantifying the similarity between two partitions that is grounded in common membership of nodes induced by the partition as opposed to overlaps between clusters \cite{gates2019element}. We use ECS\footnote{For computing ECS, we use the value of $0.9$ for the $\alpha$ parameter as suggested in \cite{gates2019element} and used in the documentation of the CluSim Python library.} because of the methodological advantages it offers compared to most commonly used metrics including the NMI, the Jaccard index, the Fowlkes-Mallows index, the adjusted Rand index, and the F measure \cite{gates2019element}. 


While the partition that maximizes modularity is often unique \cite{aref2023suboptimality}, some graphs have multiple optimal partitions. We obtain all optimal partitions of the networks using the Gurobi solver by running it with a special configuration for finding all optimal partitions \cite{gurobi}. In cases of networks with multiple optimal partitions, we calculate AMI, RMI, and ECS for the partition of each algorithm and each of the multiple globally optimal partitions of that graph. We then conservatively report the maximum AMI, maximum RMI, and maximum ECS of each algorithm on that network to quantify the similarity between that partition and its closest optimal partition. Consequently, a low value of AMI, RMI, or ECS reported for a partition indicates its dissimilarity to any optimal partition of that network.

\subsection{Illustrative example}

\begin{figure}[ht!]
    \centering
    \includegraphics[width=0.3\textwidth]{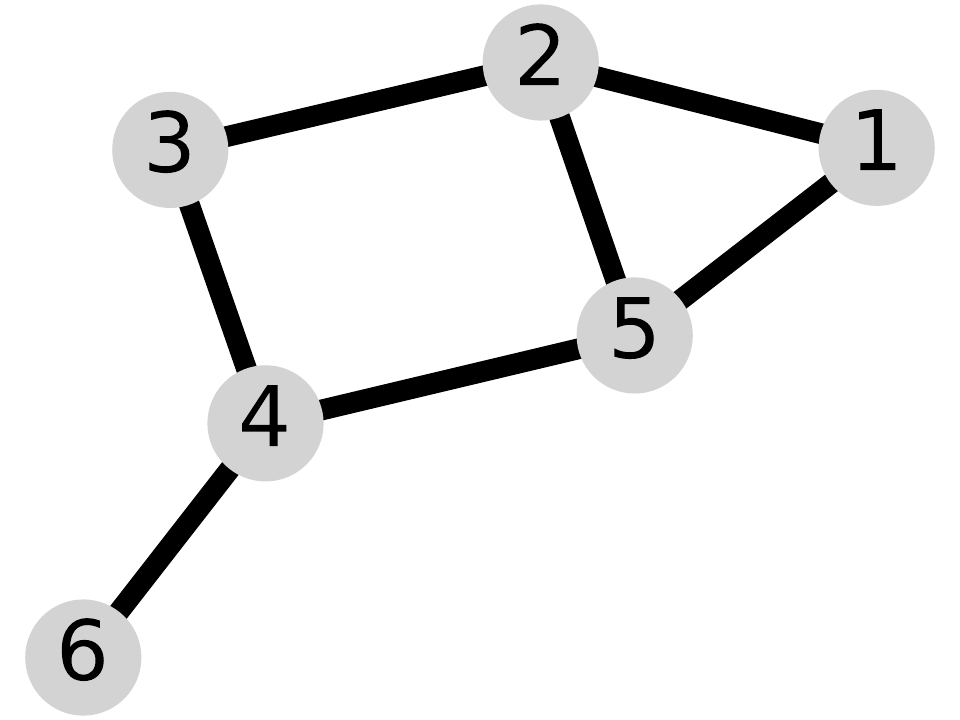}
    \hfill
    \includegraphics[width=1\textwidth]{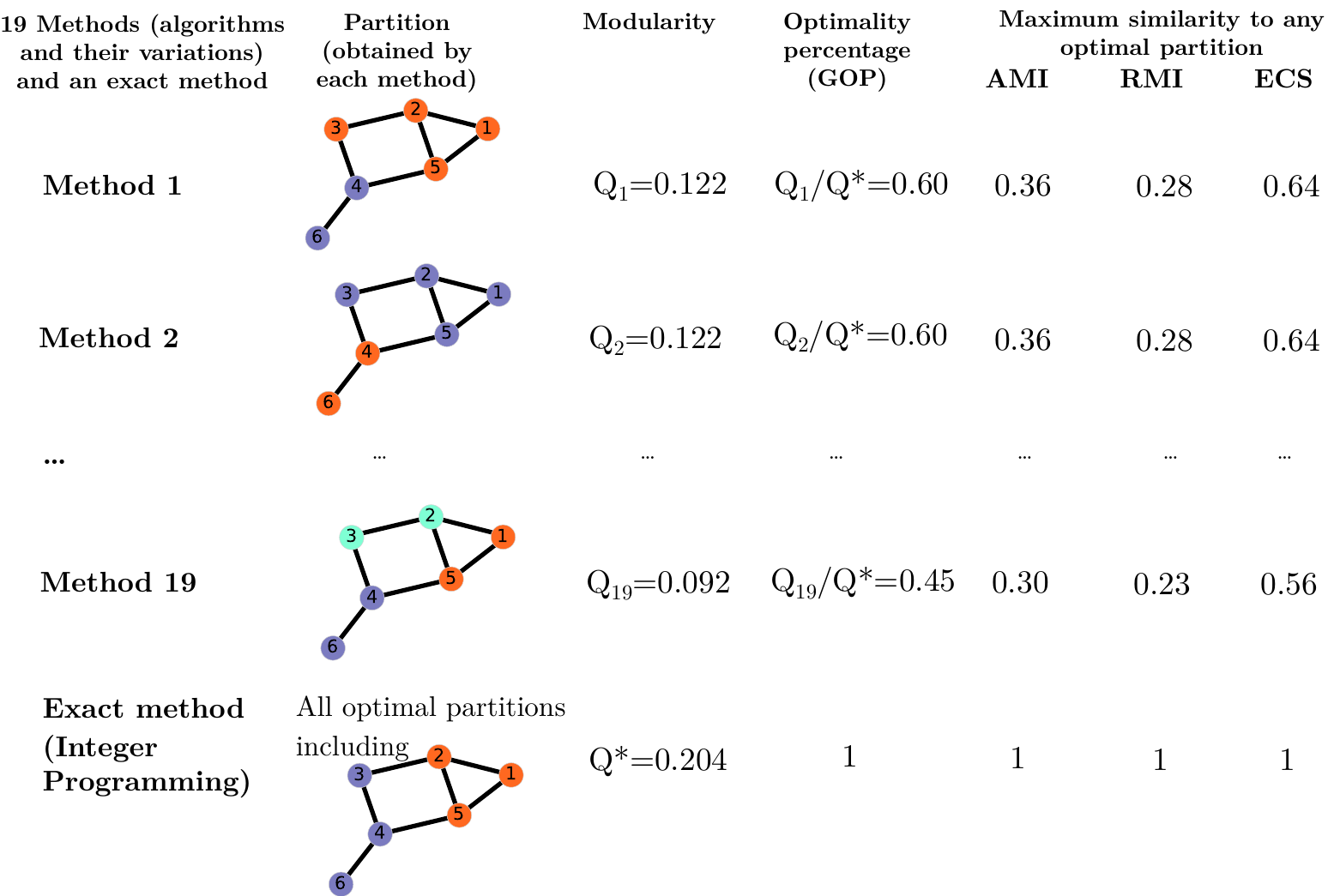}
    \caption{A toy example demonstrating several sub-optimal and one optimal partition for a graph and the corresponding modularity, GOP, AMI, RMI, and ECS values. All information and values are related to the graph shown on top.}
    \label{fig:toy}
\end{figure}

Figure~\ref{fig:toy} shows a toy example of one graph partitioned by 19 different method to demonstrate sub-optimal and optimal partitions as well as values taken by modularity, GOP, AMI, RMI, and ECS. The graph shown in Figure~\ref{fig:toy} has six nodes and seven edges. In our analysis, each network instance is partitioned by an exact method as well as ten modularity-based algorithms and their variations (19 inexact methods). 

The first row in Figure~\ref{fig:toy} shows the partition and values pertaining to method 1. Method 1 produces the partition [[1,2,3,5],[4,6]] in a (failed) attempt to maximize modularity for the input graph. The modularity of this sub-optimal partition equals $Q_1=0.122$. The same graph is also partitioned by solving the exact IP model. This method returns all optimal partitions of the input graph. In case of this graph, there is only one optimal partition which is [[1,2,5],[3,4,6]]. The modularity of this optimal partition equals $Q^*=0.204$ which is the maximum value that the modularity function can possibly take for this input graph. 

By comparing the results from method 1 and the exact method, four measures are calculated for method 1: GOP, AMI, RMI, and ECS. GOP for the partition obtained by method 1 equals $0.122/0.204=0.60$ which means the partition for method 1 is within 1-GOP$=40\%$ of the maximum modularity. The AMI between the optimal partition and the partition obtained by method 1 equals $0.36$. RMI and ECS are also calculated which reflect other perspectives on the extent of similarity of this sub-optimal partition to the optimal partition.

Note that for a any inexact method that happens to return an optimal partition for a graph, GOP, AMI, RMI, and ECS will all take the value 1 which is the desired value for all four measures. In a practical setting, AMI, RMI, and ECS (as defined in this paper) are only available when an optimal partition is available which is not always the case. Like the differences between partitions and values for Method 1 and Method 19 in Figure~\ref{fig:toy}, Our comparative analysis reveals the differences between the performance of algorithms w.r.t maximizing modularity. Method 1 has relatively more success than Method 19 according to GOP, AMI, RMI, and ECS. This is consistent with the intuitions from visually inspecting the partitions in Figure~\ref{fig:toy}.

\subsection{Specifications of network data and computing resources}
\label{ss:data}

For our computational experiments, we consider {54 real networks}\footnote{The 54 real networks are accessible from the \href{https://networks.skewed.de/}{Netzschleuder} with the details in the Appendix.} from a wide range of contexts and domains including online and offline social relations, social affiliation, social collaboration, animal interactions, biological, neural, informational, technological, fictional, geographical, organizational, communications, and terrorism. We also use 50 structurally diverse synthetic networks that have modular structures.

To create synthetic graphs with modular structures, we use two benchmark graph generation models: \textit{Lancichinetti-Fortunato-Radicchi} (LFR) benchmark graphs \cite{lancichinetti_benchmark_2008} and \textit{Artificial Benchmarks for {Community} {Detection}} (ABCD) graphs \cite{kaminski_artificial_2021}.
These two families of synthetic benchmark graphs are designed for evaluating the performance of CD algorithms based on their success in retrieving a planted (ground-truth) partition (see \cite{khomami_new_2018,singh_disintegrating_2022,aref2022bayan} for the common use case). However, we deploy these two models for generating synthetic graphs with controllable modular structures and do not use the planted partition information. LFR and ABCD each has a distinct mixing parameter which controls the association between the structure and the planted communities. This association in turn impacts the strength of the modular structure (relatively higher density of intra-community edges compared to the density of inter-community edges). ABCD is the more recent alternative to the LFR model and offers additional benefits including higher scalability and better control for adjusting the mixing parameter \cite{kaminski_artificial_2021}. 

The 50 synthetic networks include 20 LFR graphs and 30 ABCD graphs with up to 1000 edges. The real networks considered have at most 2812 edges. These instance sizes were chosen to ensure that globally optimal partitions can be obtained in a reasonable time. Publicly available links and additional details for both real and synthetic networks are provided in the appendix.

The computational experiments were implemented in Python 3.9 using a notebook computer with an Intel Core i7-11800H @ 2.30GHz CPU and 64 GB of RAM running Windows 10.

\section{Results}
\label{s:results}

We present the main results from our experiments in the following five subsections. In Subsection \ref{ss:facebook}, we compare partitions from 12 modularity maximization methods on a single network to illustrate the differences between different methods (algorithms and their variations) for solving the same underlying optimization problem. In Subsection \ref{ss:scatter}, we use AMI, RMI, and ECS to investigate the cost of sub-optimality in terms of dissimilarity of partitions from an optimal partition. In Subsection \ref{ss:similarity}, we provide the distributions of AMI, RMI, and ECS for each algorithm on all 104 networks. In Subsection \ref{ss:time}, we compare the solve times of all the algorithms and their variations. Finally, in Subsection \ref{ss:success}, we investigate the success rate of all the algorithms and their variations in returning an optimal partition.

\subsection{Comparing partitions from different algorithms on one network}\label{ss:facebook}
\begin{figure}[!htbp]
     \centering
          \begin{subfigure}[b]{0.385\textwidth}
         \centering
         \includegraphics[trim={8.6cm 8.6cm 8.6cm 8.6cm},clip,width=\textwidth]{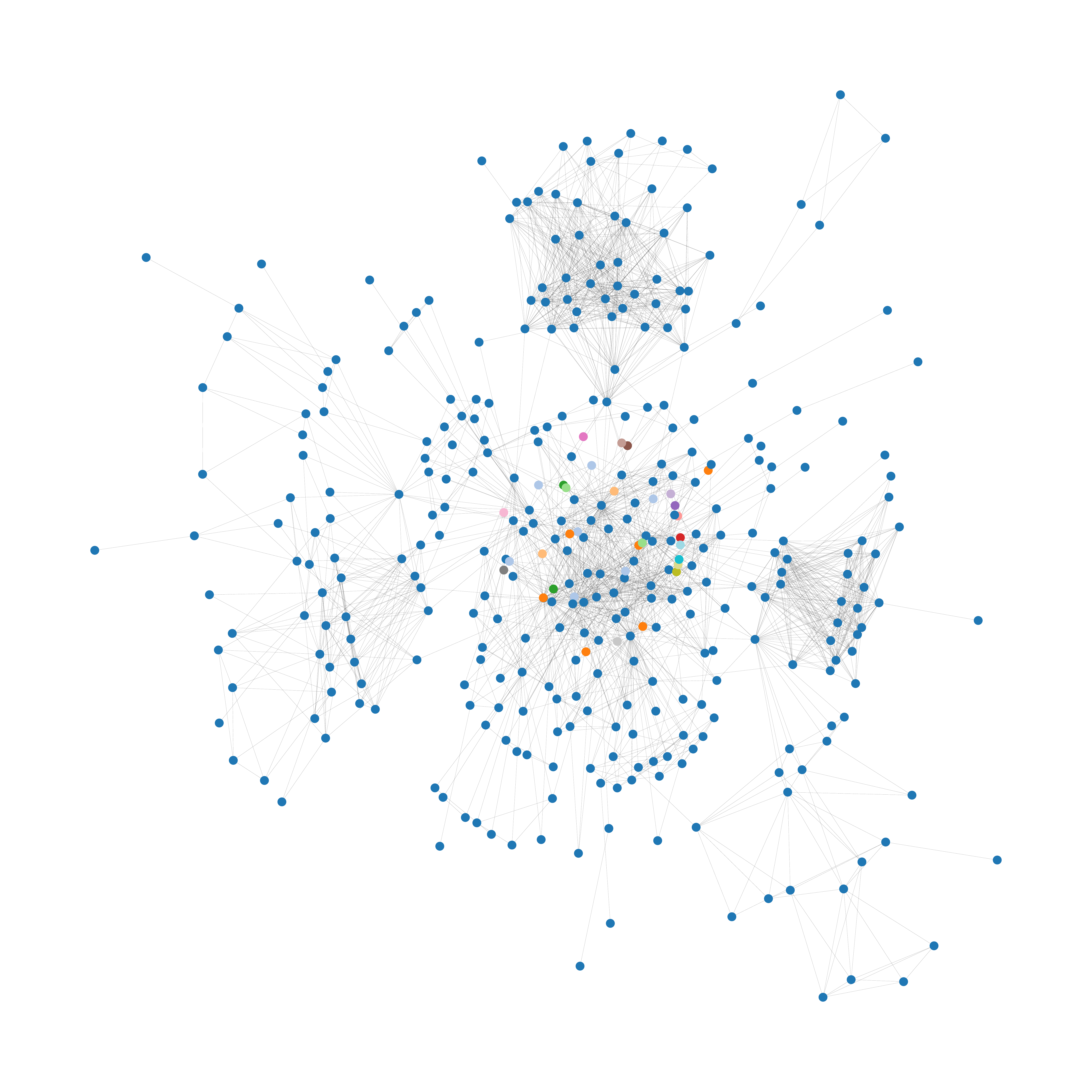}
         \caption{Paris, $Q=.0338,\\\hspace{\textwidth} k=20, \text{AMI}=.363$}
         \label{subfig:paris}
     \end{subfigure}
    \begin{subfigure}[b]{0.385\textwidth}
         \centering
         \includegraphics[trim={8.6cm 8.6cm 8.6cm 8.6cm},clip,width=\textwidth]{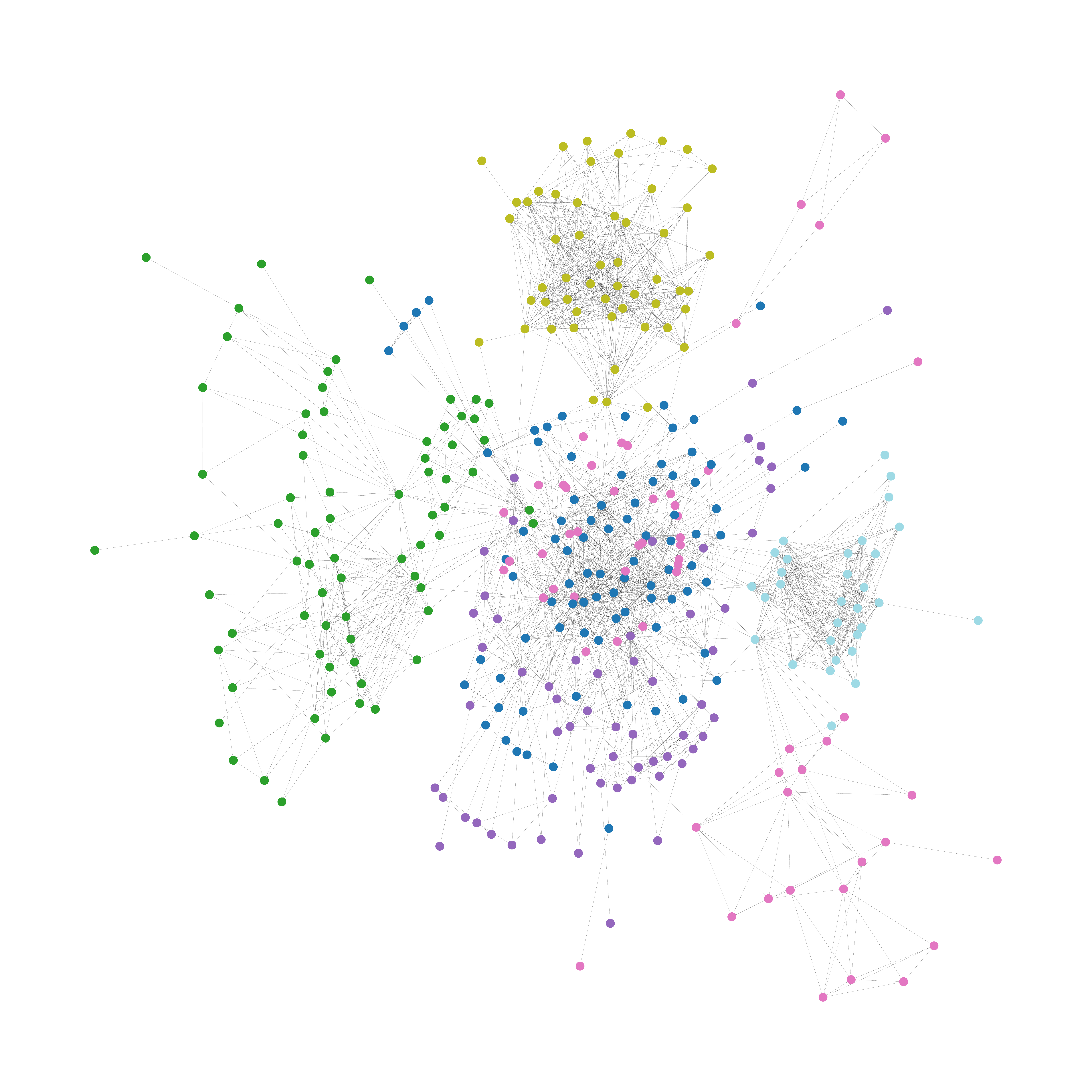}
         \caption{Belief, $Q=.4566, \\\hspace{\textwidth} k=6, \text{AMI}=.786$}
         \label{subfig:belief}
     \end{subfigure}
     \begin{subfigure}[b]{0.385\textwidth}
         \centering
         \includegraphics[trim={8.6cm 8.6cm 8.6cm 8.6cm},clip,width=\textwidth]{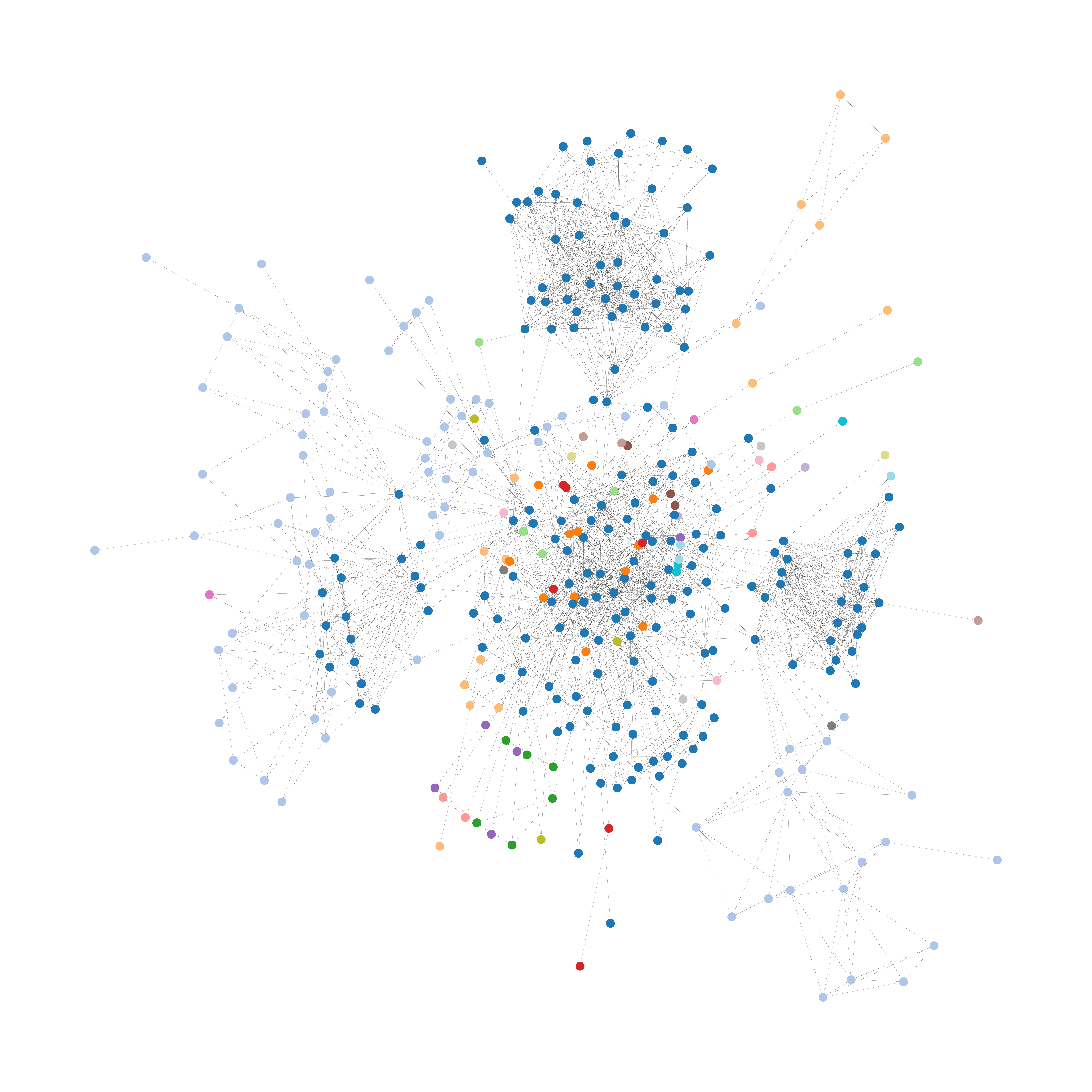}
         \caption{EdMot, $Q=.4902,\\\hspace{\textwidth} k=53, \text{AMI}=.651$}
         \label{subfig:edmot}
     \end{subfigure}
    \begin{subfigure}[b]{0.385\textwidth}
         \centering
         \includegraphics[trim={8.6cm 8.6cm 8.6cm 8.6cm},clip,width=\textwidth]{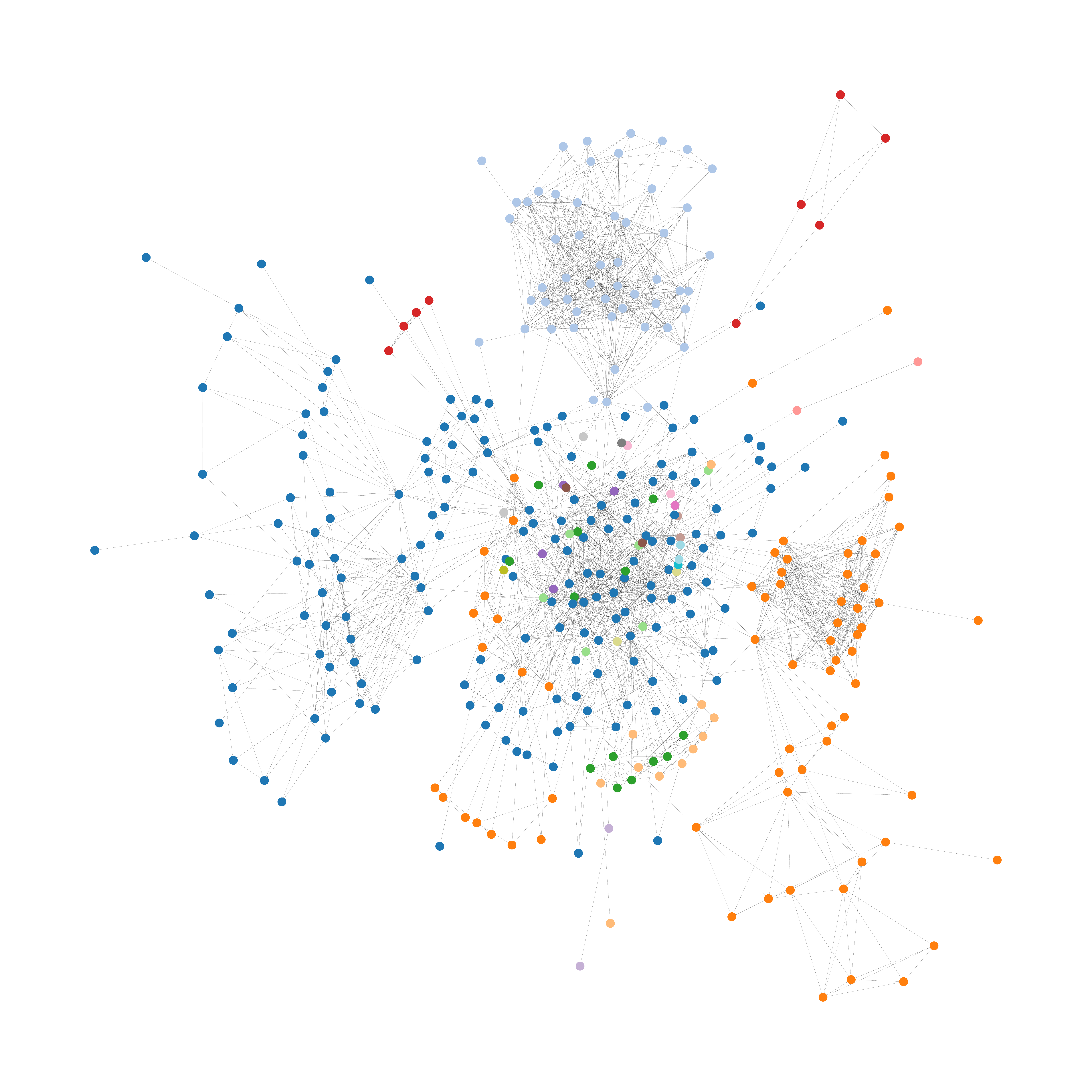}
         \caption{CNM, $Q=.6971,\\\hspace{\textwidth} k=30, \text{AMI}=.829$}
         \label{subfig:cnm}
     \end{subfigure}
     \begin{subfigure}[b]{0.385\textwidth}
         \centering
         \includegraphics[trim={8.6cm 8.6cm 8.6cm 8.6cm},clip,width=\textwidth]{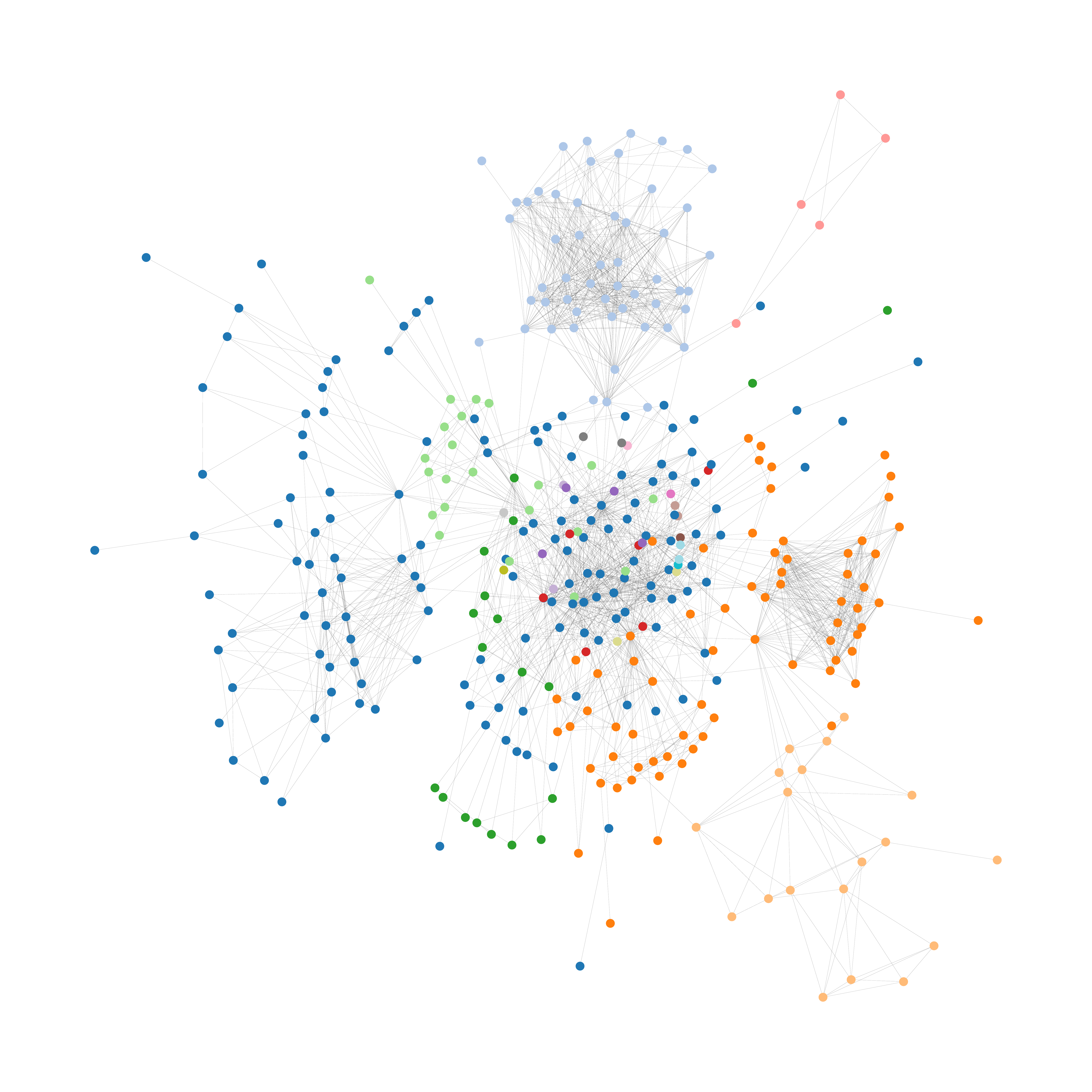}
         \caption{Leiden, $Q=.7082,\\\hspace{\textwidth} k=32, \text{AMI}=.908$}
         \label{subfig:leiden}
     \end{subfigure}
     \begin{subfigure}[b]{0.385\textwidth}
         \centering
         \includegraphics[trim={8.6cm 8.6cm 8.6cm 8.6cm},clip,width=\textwidth]{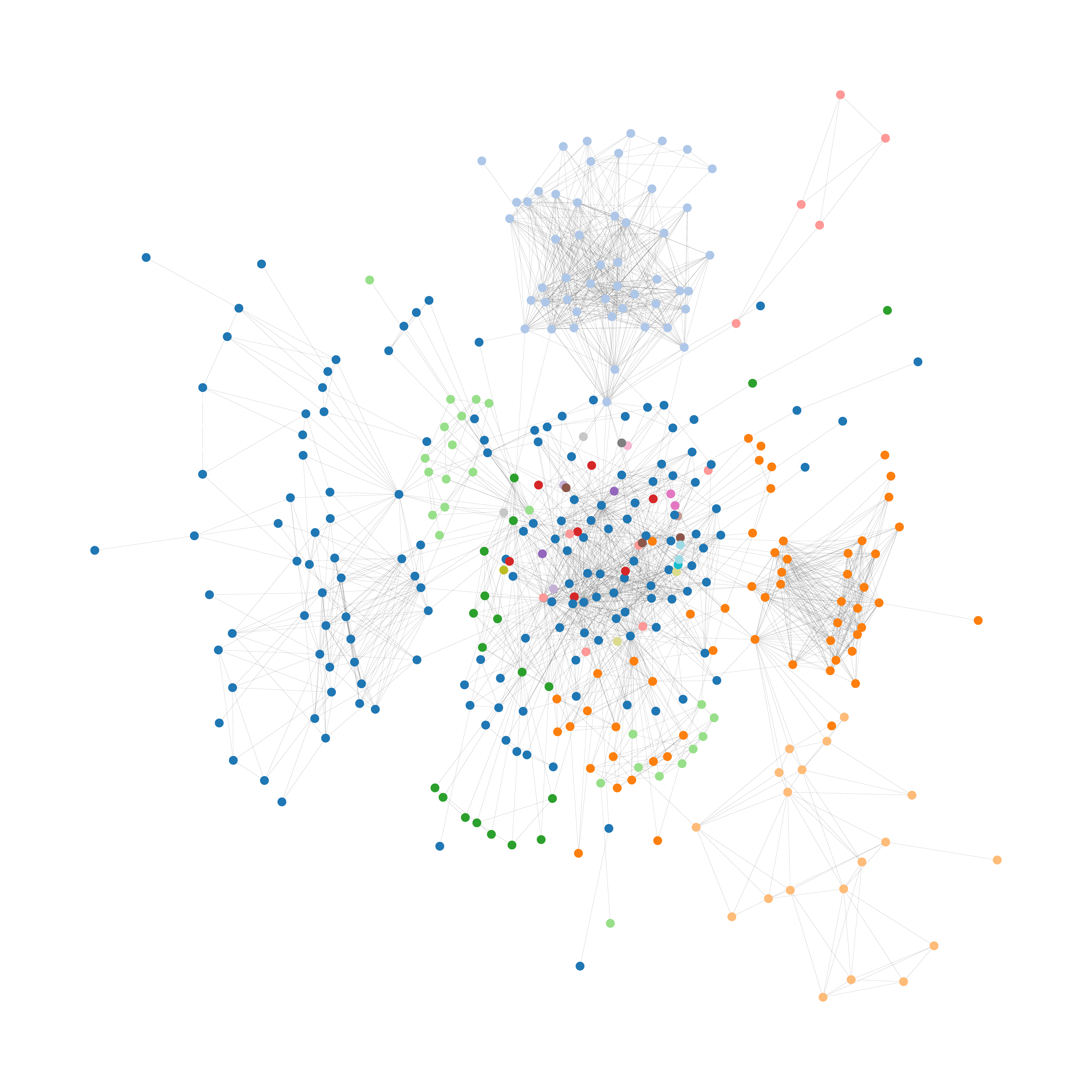}
         \caption{Louvain, $Q=.7087,\\\hspace{\textwidth} k=29, \text{AMI}=.920$}
         \label{subfig:louvain}
     \end{subfigure}
        \caption{Modularity maximization for one network using six methods leading to six sub-optimal partitions (panels a-f) with increasing $Q$, different $k$, and different AMI values. Only the giant component is shown. (Magnify the high-resolution color figure on screen for more details.) }
        \label{fig:facebook1}
\end{figure}

\begin{figure}[!htbp]
     \centering
    \begin{subfigure}[b]{0.385\textwidth}
         \centering
         \includegraphics[trim={8.6cm 8.6cm 8.6cm 8.6cm},clip,width=\textwidth]{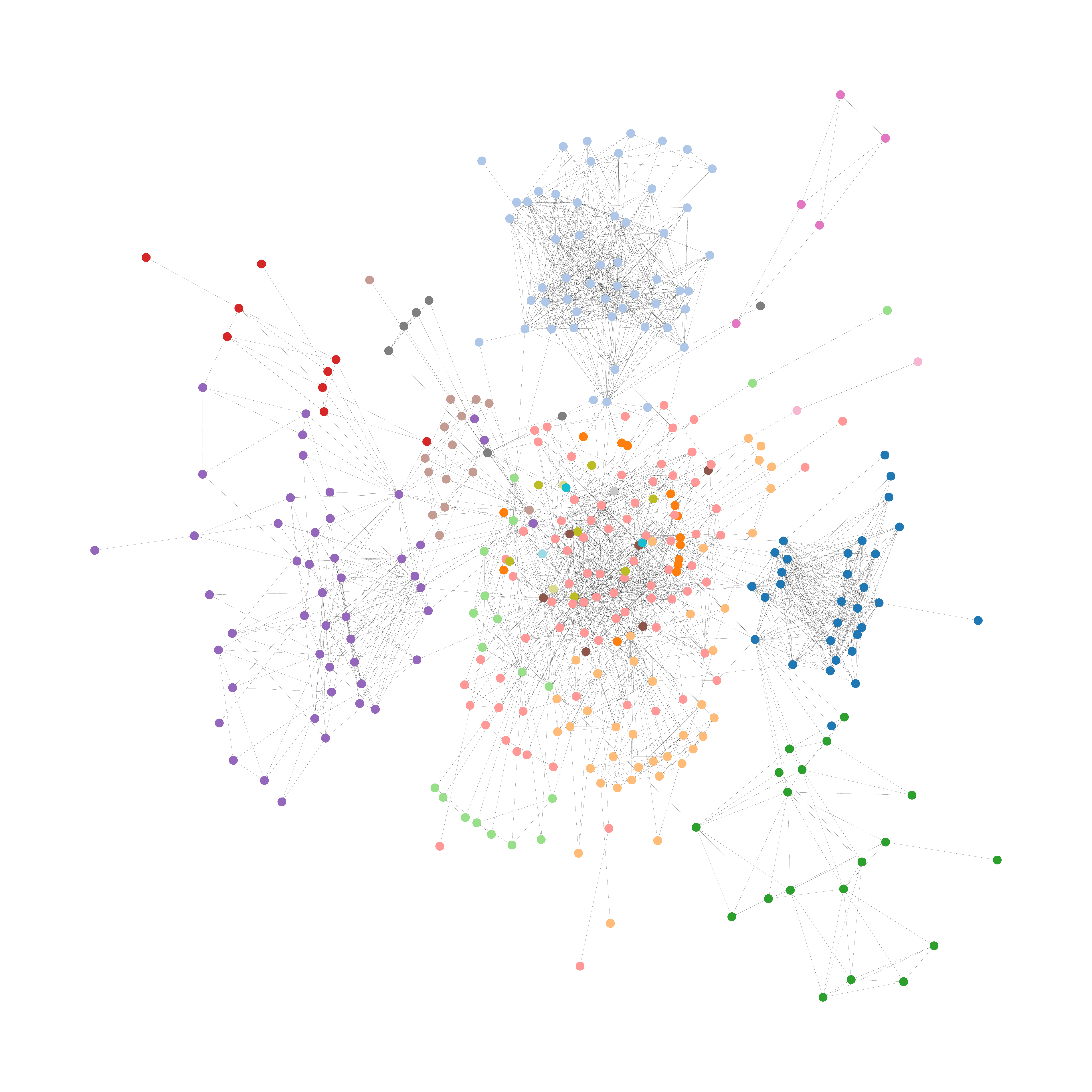}
         \caption{GNN-100, $Q=.713294,
         \\\hspace{\textwidth} k=19, \text{AMI}=.941$}
         \label{subfig:GNN-100}
     \end{subfigure}
     \begin{subfigure}[b]{0.385\textwidth}
         \centering
         \includegraphics[trim={8.6cm 8.6cm 8.6cm 8.6cm},clip,width=\textwidth]{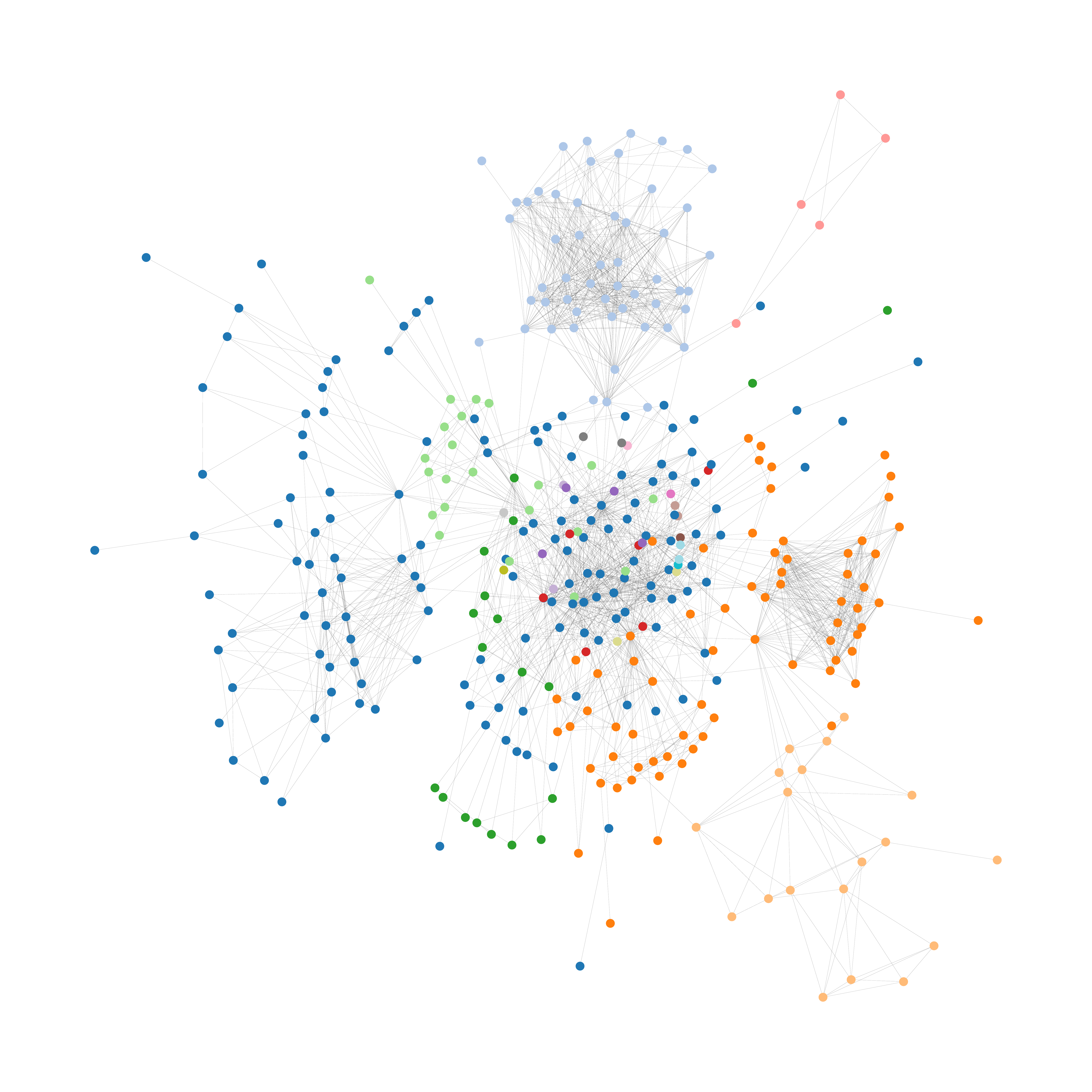}
         \caption{LN, $Q=.7153755,\\\hspace{\textwidth} k=28, \text{AMI}=.971$}
         \label{subfig:ln}
     \end{subfigure}
    \begin{subfigure}[b]{0.385\textwidth}
         \centering
         \includegraphics[trim={8.6cm 8.6cm 8.6cm 8.6cm},clip,width=\textwidth]{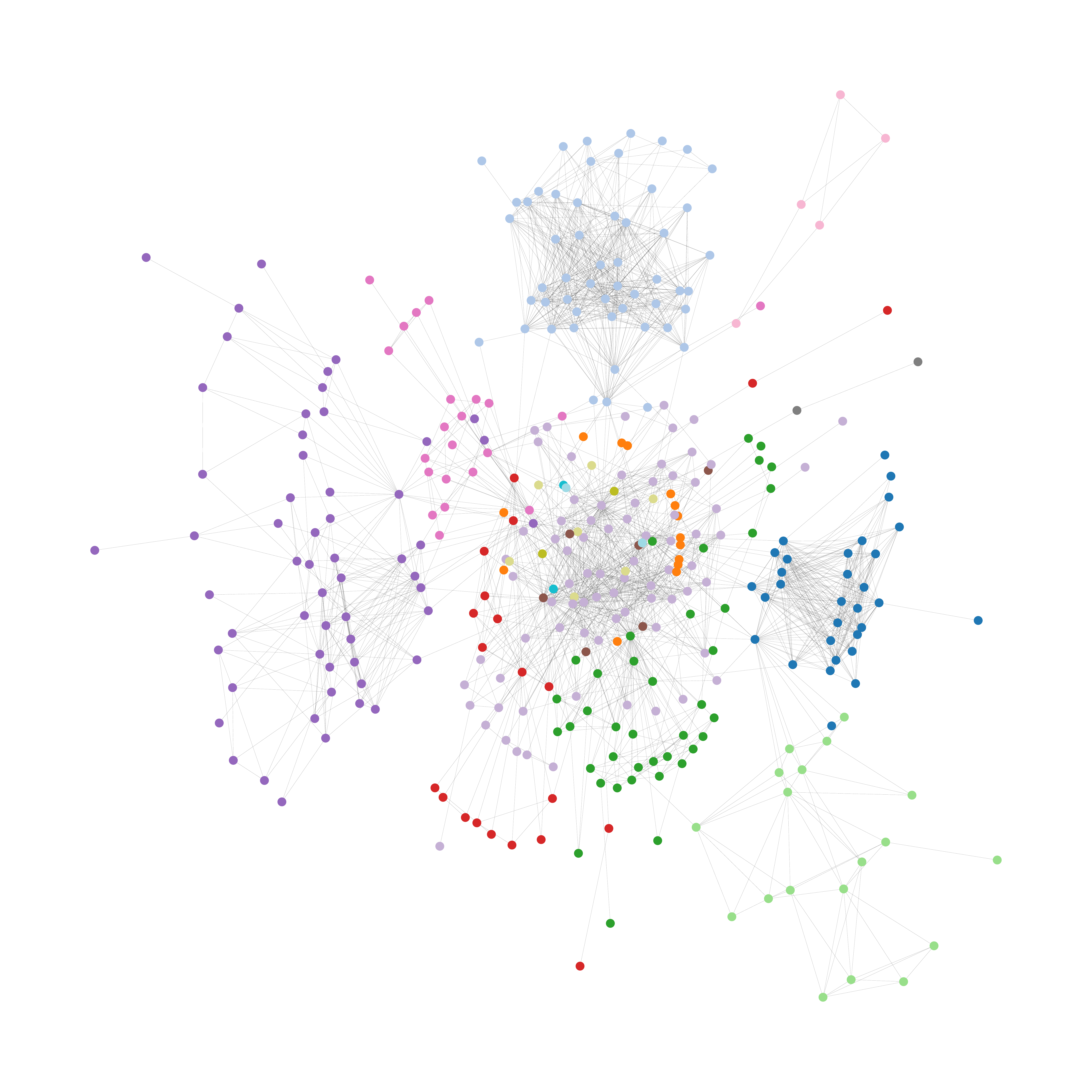}
         \caption{GNN-25K, $Q=.715628, \\\hspace{\textwidth} k=16, \text{AMI}=.957$}
         \label{subfig:gnn25000}
     \end{subfigure}
     \begin{subfigure}[b]{0.385\textwidth}
         \centering
         \includegraphics[trim={8.6cm 8.6cm 8.6cm 8.6cm},clip,width=\textwidth]{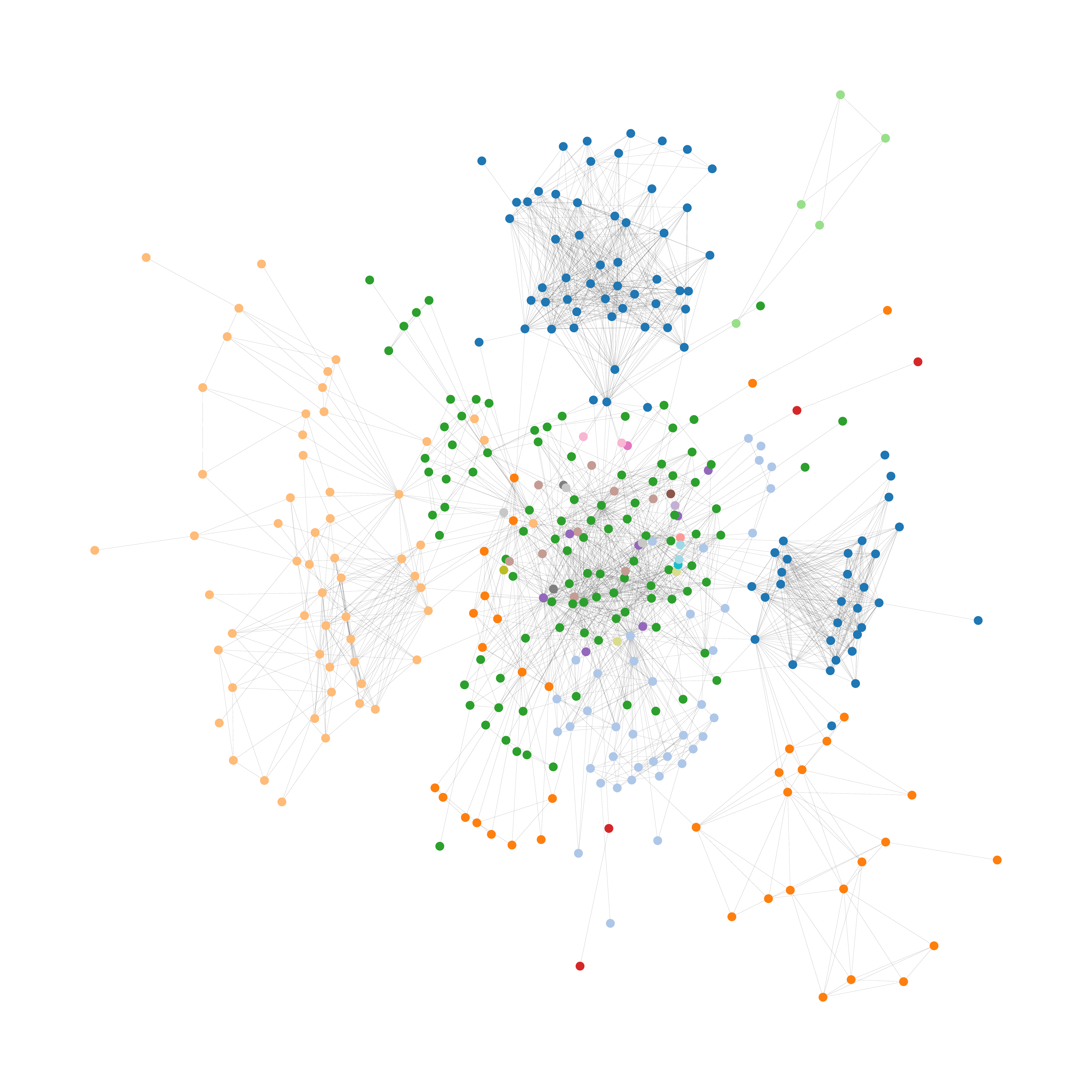}
         \caption{Bayan 0.9, $Q=.7156751,\\\hspace{\textwidth} k=30, \text{AMI}=.987$}
         \label{subfig:bayan0.9}
     \end{subfigure}
               \begin{subfigure}[b]{0.385\textwidth}
         \centering
         \includegraphics[trim={8.6cm 8.6cm 8.6cm 8.6cm},clip,width=\textwidth]{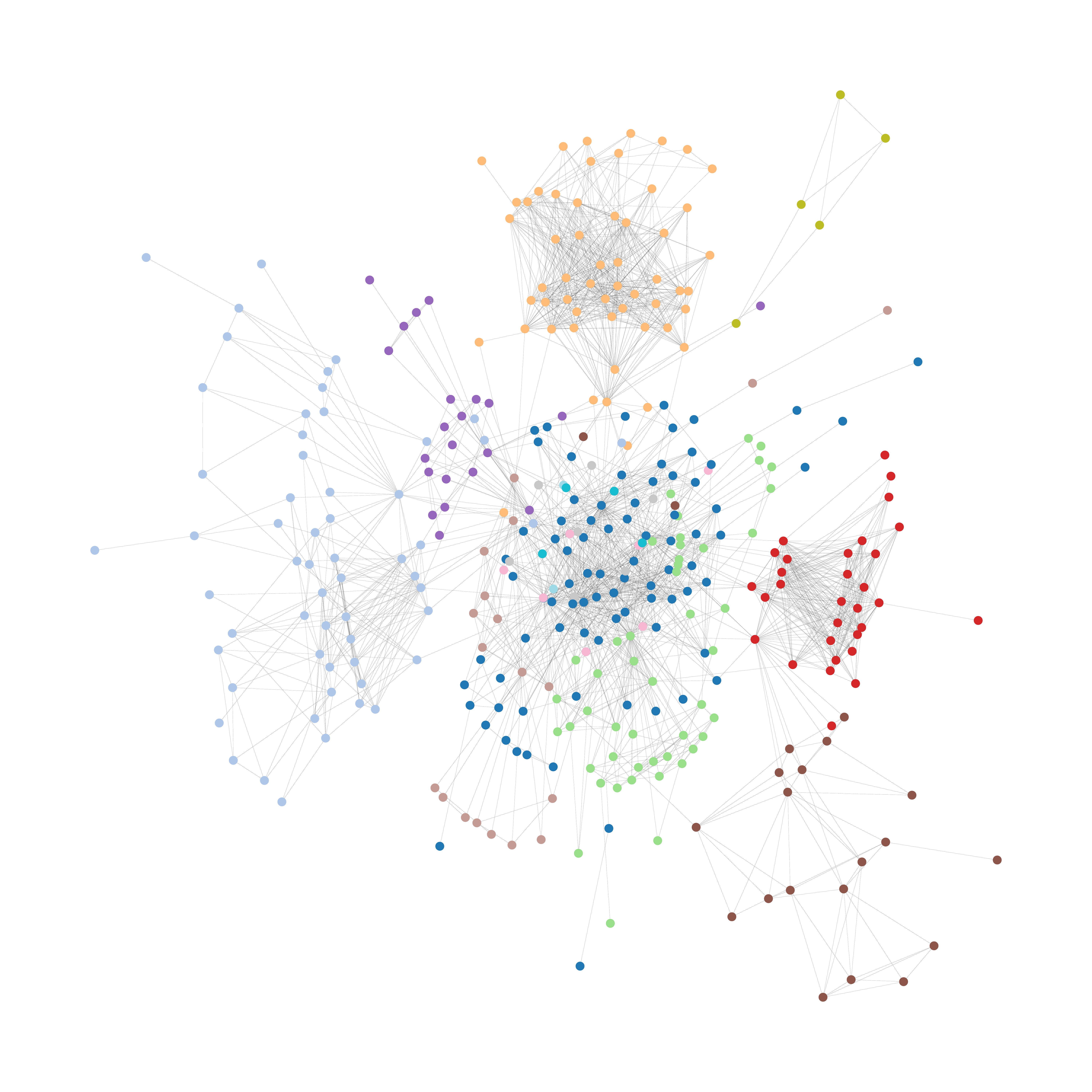}
         \caption{Combo, $Q=.7157709,\\\hspace{\textwidth} k=13, \text{AMI}=.949$}
         \label{subfig:combo}
     \end{subfigure}
     \begin{subfigure}[b]{0.385\textwidth}
         \centering
         \includegraphics[trim={8.6cm 8.6cm 8.6cm 8.6cm},clip,width=\textwidth]{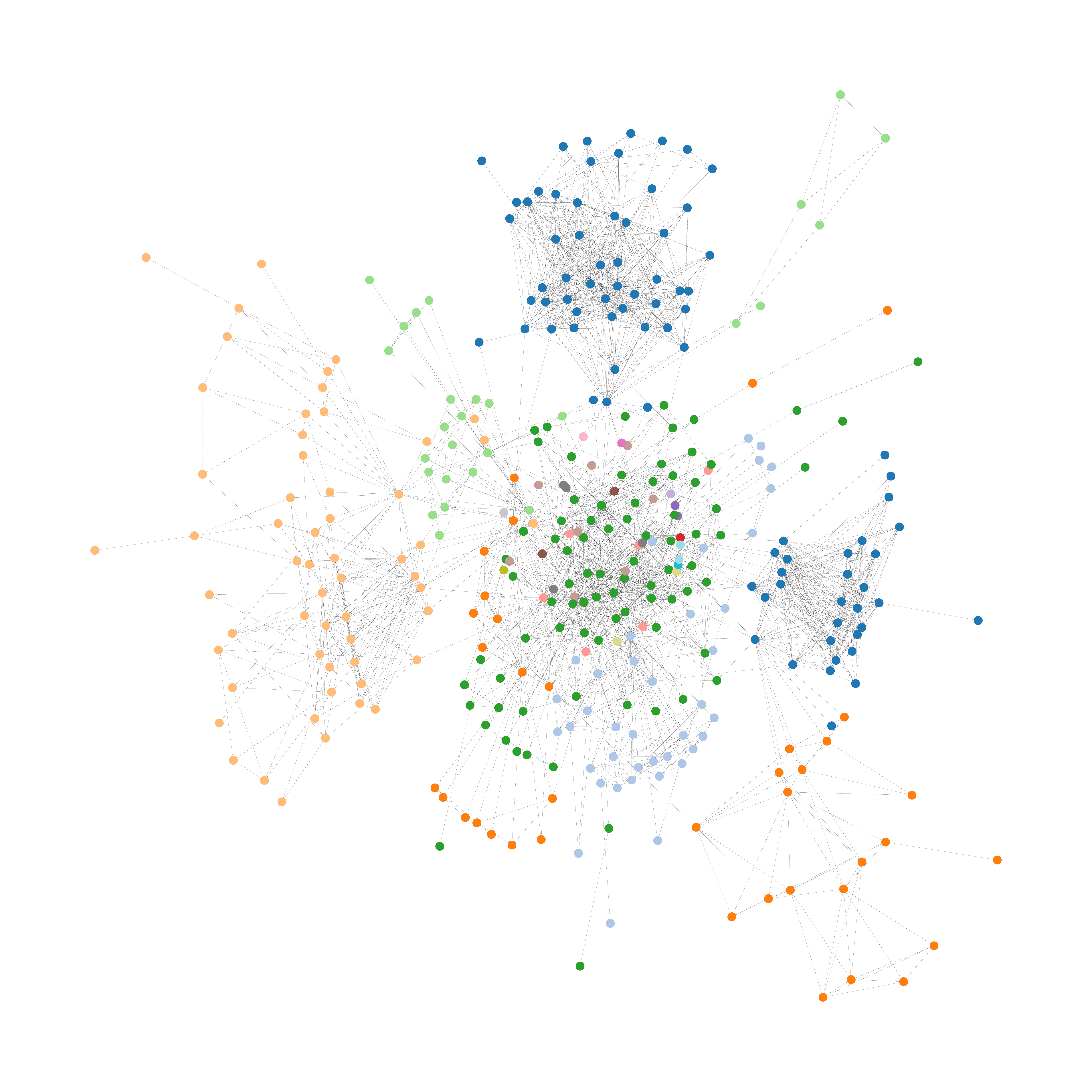}
         \caption{Bayan 1e-3, $Q^*=.7157714,\\\hspace{\textwidth} k=28, \text{AMI}=1$}
         \label{subfig:bayan1e-3}
     \end{subfigure}
        \caption{Modularity maximization for one network using six methods leading to one optimal partition (panel f) and five sub-optimal partitions (panels a-e) with increasing $Q$, different $k$, and different AMI values. Only the giant component is shown. (Magnify the high-resolution color figure on screen for more details.) }
        \label{fig:facebook2}
\end{figure}

Figures~\ref{fig:facebook1}--\ref{fig:facebook2} show the largest connected component (the giant component) of one network that is partitioned by twelve modularity-based CD methods. This network\footnote{ \emph{facebook\_friends} network \cite{maier2017cover} from the \href{https://networks.skewed.de/}{Netzschleuder} repository} represents an anonymized Facebook \textit{egocentric network}\footnote{A network of one person's social ties to other persons and their ties to each other} from which the ego node has been removed. Nodes represent Facebook users, and an edge exists between any pair of users who were friends on Facebook in April 2014 \cite{maier2017cover}. Partitions of nodes into communities are shown using node colors.

Comparing Figures~\ref{fig:facebook1}--\ref{fig:facebook2}, the partitions from the six algorithms in Figure~\ref{fig:facebook1} have more substantial differences from the optimal values in $Q$, AMI, and $k$ (number of communities) as shown by the values in the corresponding subcaptions in Figure~\ref{fig:facebook1}. Subfigure~\ref{subfig:bayan1e-3} shows an optimal partition of the network obtained using the Bayan approximate algorithm with an approximation tolerance of $1e-3$. This optimal partition has $k=28$ communities, and a modularity value of $Q^*=0.7157714$ (the maximum modularity for this network). The partitions from the all the other eleven methods are sub-optimal. Compared to other heuristics, the two heuristics Combo and LN have more success in achieving proximity to an optimal partition. LN returns a partition with $k=28$ communities and a modularity of $Q=0.7153755$ which has the highest AMI among all heuristics ($0.971$). The relative success of the Combo algorithm is in returning a particularly high-modularity partition with $Q=0.7157709$, but with $k=13$ communities and a lower AMI ($0.949$) compared to the AMI of LN. The two variations of the GNN algorithm return sub-optimal partitions with 19 and 16 communities. A similar observation can be made from the RMI and ECS values of these partitions which are not reported in the interest of brevity.

\FloatBarrier

\subsection{The disproportionate cost of sub-optimality}
\label{ss:scatter}

\begin{figure}[!hbp]
\centering
	\includegraphics[width=1\textwidth]{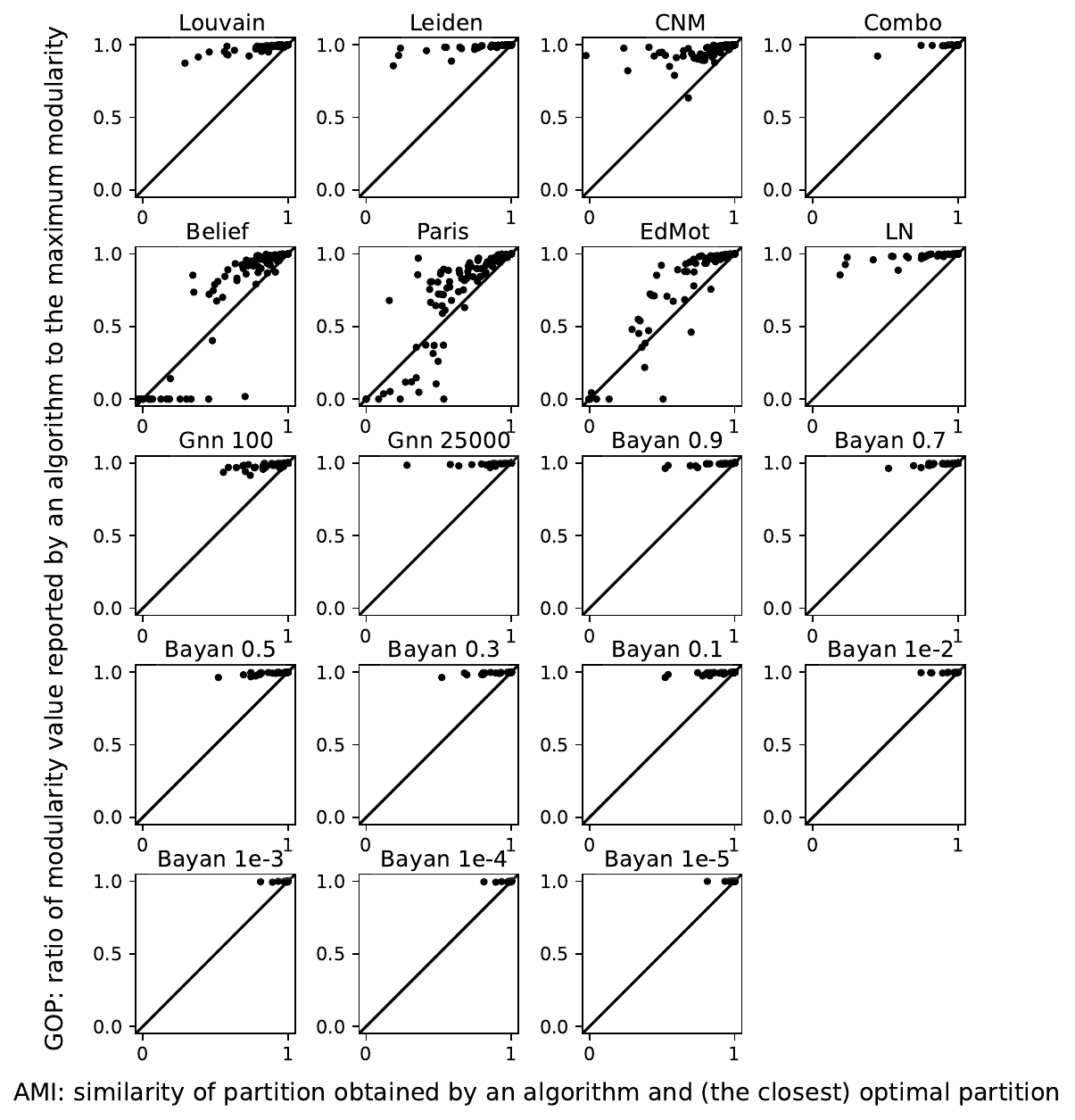}
	\caption{Global optimality percentage (GOP) and normalized adjusted mutual information (AMI) measured for each algorithm by comparing its results with (all) globally optimal partitions. (Magnify the high-resolution figure on screen for more details.)}
	\label{fig:GOP_AMI}
\end{figure}

\begin{figure}[!htbp]
\centering
	\includegraphics[width=1\textwidth]{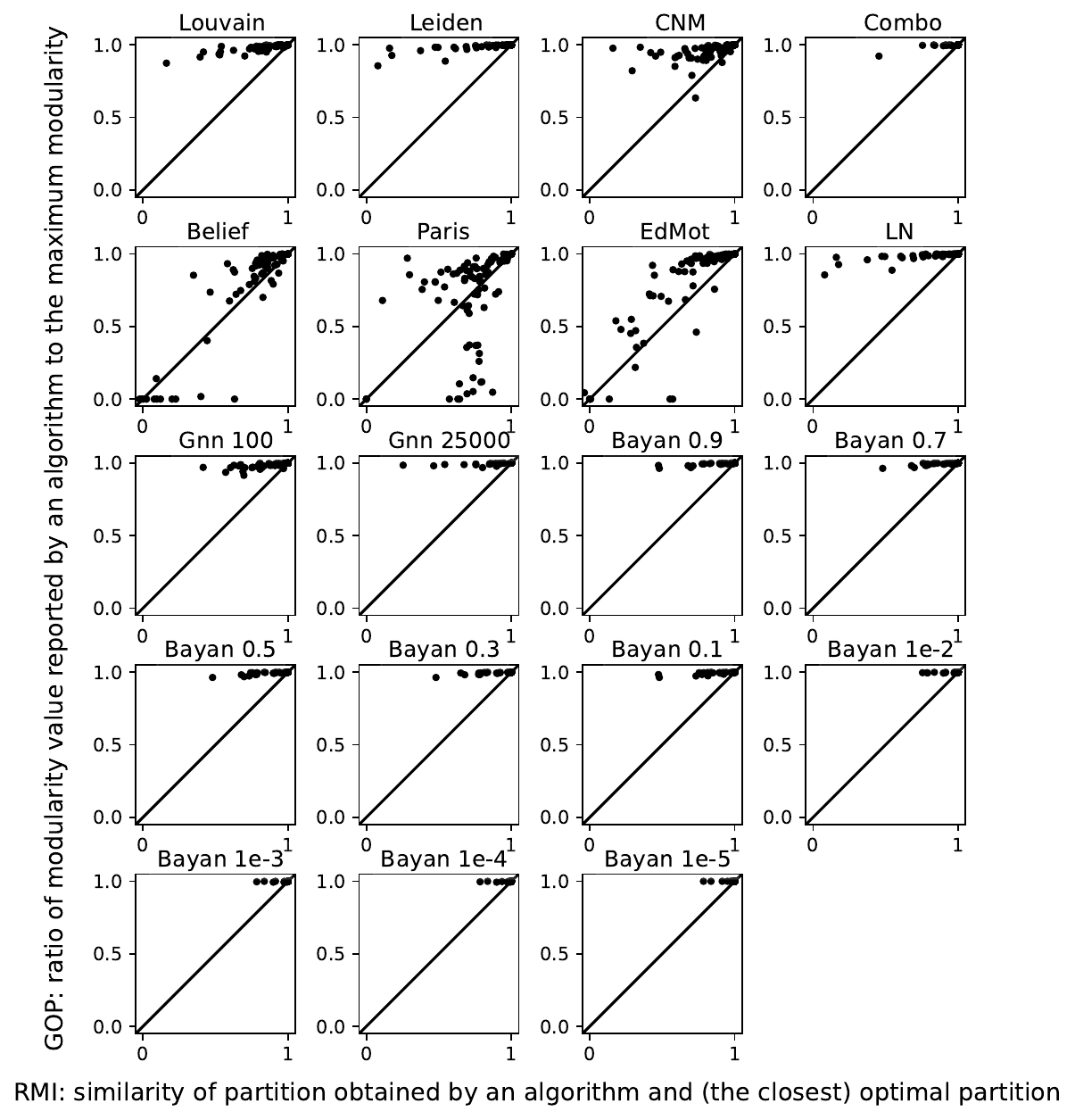}
	\caption{Global optimality percentage (GOP) and normalized reduced mutual information (RMI) measured for each algorithm by comparing its results with (all) globally optimal partitions. (Magnify the high-resolution figure on screen for more details.)}
	\label{fig:GOP_RMI}
\end{figure}

\begin{figure}[!htbp]
\centering
	\includegraphics[width=1\textwidth]{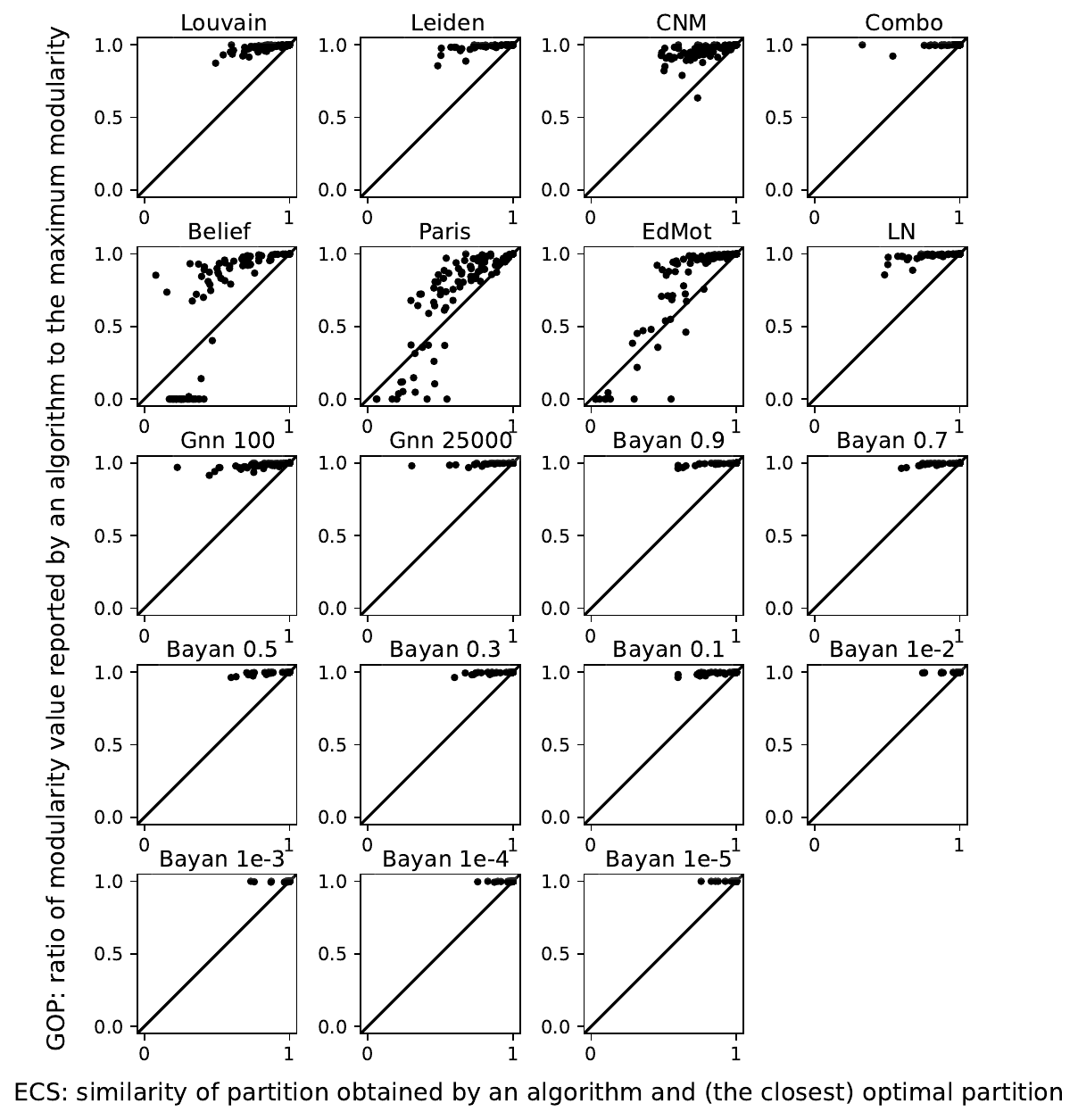}
	\caption{Global optimality percentage (GOP) and element-centric similarity (ECS) measured for each algorithm by comparing its results with (all) globally optimal partitions. (Magnify the high-resolution figure on screen for more details.)}
	\label{fig:GOP_ECS}
\end{figure}

We use three scatter-plots in Figures~\ref{fig:GOP_AMI}--\ref{fig:GOP_ECS} to investigate the cost of sub-optimality from the three perspectives of AMI, RMI, and ECS (in terms of dissimilarity from an optimal partition) for each of the algorithms based on all the 104 networks. 

Figure~\ref{fig:GOP_AMI} has one panel for each algorithm which shows GOP on the y-axis and AMI on the x-axis. Each of the 104 data points in a panel of Figure~\ref{fig:GOP_AMI} corresponds to one of the 104 networks. The 45-degree lines in Figure~\ref{fig:GOP_AMI} indicate the areas where the GOP and AMI are equal. In other words, the 45-degree lines show areas where the extent of sub-optimality (1-GOP) is associated with a dissimilarity (1-AMI) of the same proportion between the sub-optimal partition and any optimal partition.

Looking at the y-axes values in Figure~\ref{fig:GOP_AMI}, we observe that there is a substantial variation in the values of GOP (i.e., variations in the extent of sub-optimality) for most of the eight heuristic algorithms. The Belief algorithm returns partitions associated with negative modularity values for 22 of the 104 instances (leading to the corresponding data points having GOP$=0$ and being concentrated at the bottom of the scatter-plot). The Paris and EdMot algorithms return partitions with modularity values substantially smaller than the maximum modularity values. Among the eight heuristics, the four algorithms with the highest and increasing performance in returning close-to-maximum modularity values are LN, Leiden, Louvain, and Combo. Despite that these instance are graphs with no more than 2812 edges, they are, according to Figure~\ref{fig:GOP_AMI}, challenging instances for these heuristic algorithms to optimize. Given that modularity maximization is an NP-hard problem \cite{brandes2007modularity,meeks2020parameterised}, one can argue that the performance of these heuristics, in achieving proximity to an optimal partition, does not improve for larger networks. The y-axes values for different variations of Bayan and GNN have a much lower variability and are closer (than partitions of most heuristics) to 1. This indicates that these two algorithms return partitions with modularity values closer to the maximum modularity values of these networks. 

The x-axes values for the heuristics in Figure~\ref{fig:GOP_AMI}, except Combo, show considerable dissimilarity (from an AMI perspective) between the sub-optimal partitions of these heuristics and any optimal partition for these 104 instances. Some sub-optimal partitions obtained by these heuristics have AMI values smaller than {$0.5$}. These are substantially different from any optimal partition. Even for the data points concentrated at the top of the scatter-plots which have $0.95<\text{GOP}<1$, we see some substantially small AMI values. They indicate that some high-modularity partitions are particularly dissimilar to any optimal partition. Compared to the other seven heuristics, Combo appears to consistently return partitions with high AMIs on a larger number of these 104 instances. The twelve panels for Combo and different variations of GNN and Bayan show fewer instances of low AMI values indicating that these three algorithms are overall more successful at returning partitions highly similar to an optimal partition. The panels for Bayan show that decreasing the approximation tolerance (gradually from $0.9$ to $1e-5$) leads to a gradual increase in the resulting AMI values. Unlike heuristics whose performance cannot be controlled through a user-specified parameter, Bayan provides the user with the flexibility to obtain approximations closer to optimal by reducing the tolerance (at the cost of additional computations).

The most important pattern in Figure~\ref{fig:GOP_AMI} is observed when we focus on the positions of data points with respect to the 45-degree lines. We observe that they are mostly located above their corresponding 45-degree line. This indicates that, irrespective of the algorithm, sub-optimal partitions tend to be disproportionately dissimilar to any optimal partition (as foreshadowed in \cite{dinh_network_2015}). This result goes against the naive viewpoint that close-to-maximum modularity partitions are also close to an optimal partition. Our results are aligned with previous concerns that modularity maximization heuristics have a high risk of algorithmic failure \cite{kawamoto2019counting} and may result in degenerate solutions far from the underlying community structure \cite{good_performance_2010}.

To ensure that the observations made are not artefacts of using AMI, we also report the RMI and ECS values of the same partitions. Figures \ref{fig:GOP_RMI}--\ref{fig:GOP_ECS} have x-axes values based on RMI and ECS respectively and show GOP on their y-axes. Similar observations can be made from the RMI and ECS values of these partitions: (1) x-axes values in Figures \ref{fig:GOP_RMI}--\ref{fig:GOP_ECS} for partitions of all heuristics except Combo show substantial dissimilarity to any optimal partitions of these networks. (2) GNN, Combo, and Bayan return partitions with higher similarity to the optimal partitions, also when RMI or ECS are used. (3) Most data point are above their corresponding 45-degree lines indicating that sub-optimal partitions tend to be disproportionately dissimilar to any optimal partition (from RMI and ECS perspectives).

\FloatBarrier

\subsection{Distribution of partition similarity measures for each algorithm} 
\label{ss:similarity}

\begin{figure}[!htbp]
\begin{subfigure}[b]{\textwidth}
    \centering
    \includegraphics[width=1\textwidth]{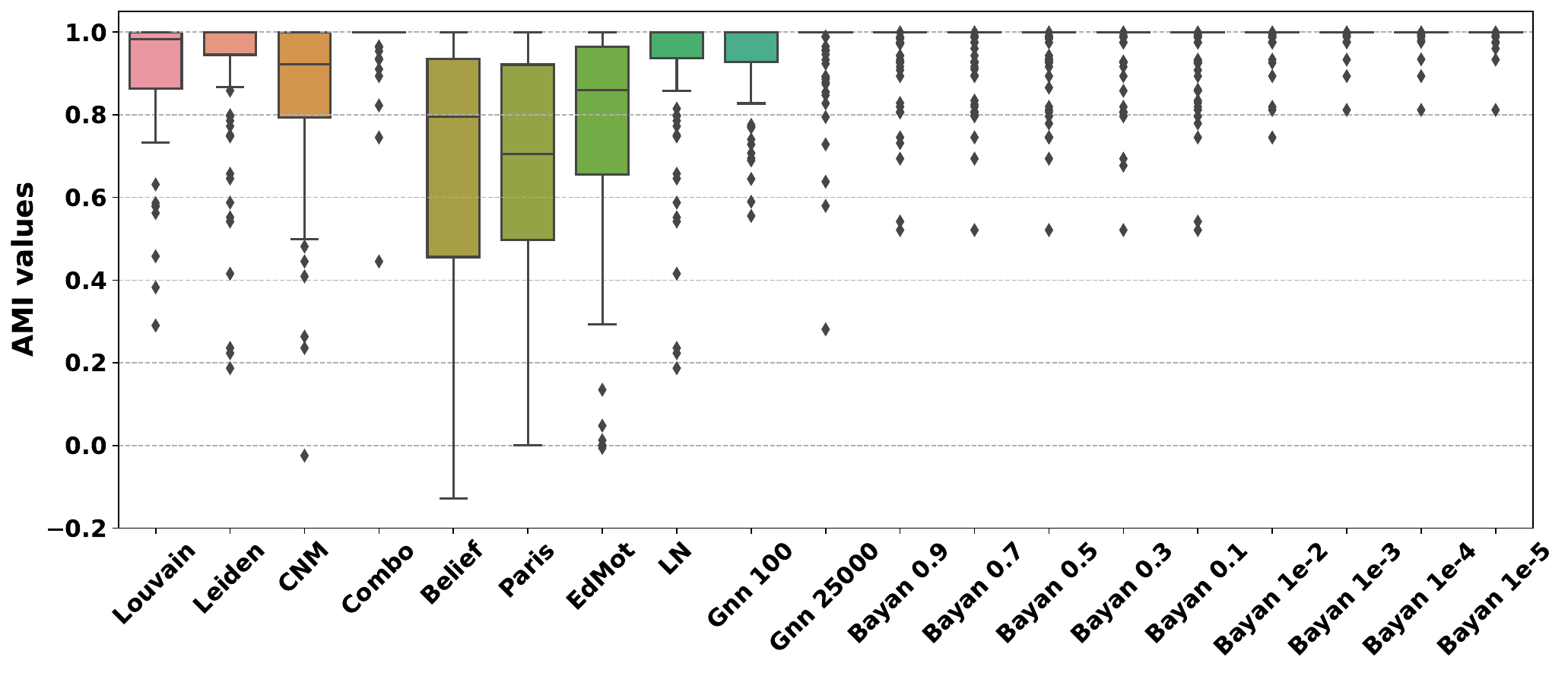}
\end{subfigure}
\begin{subfigure}[b]{\textwidth}
    \centering
	\includegraphics[width=1\textwidth]{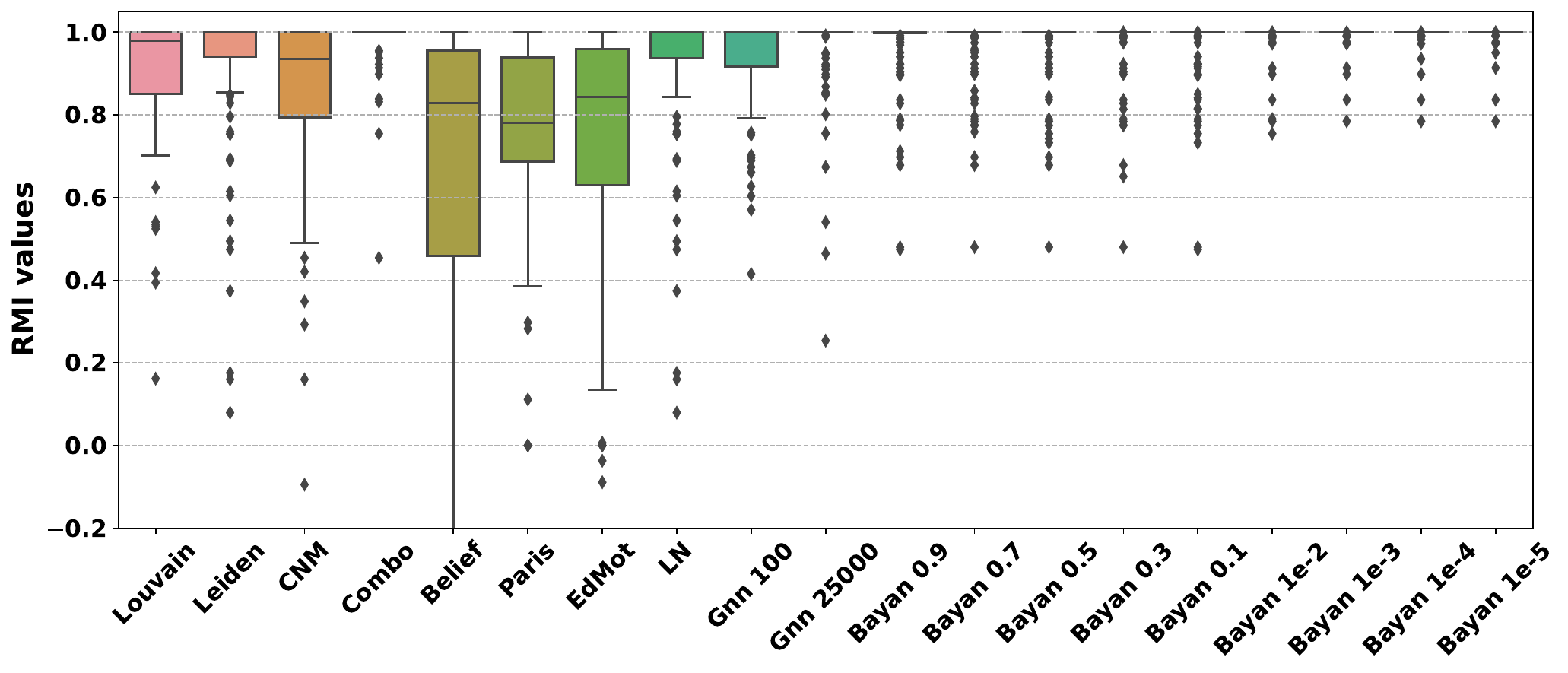}
 \end{subfigure}
\begin{subfigure}[b]{\textwidth}
    \centering
	\includegraphics[width=1\textwidth]{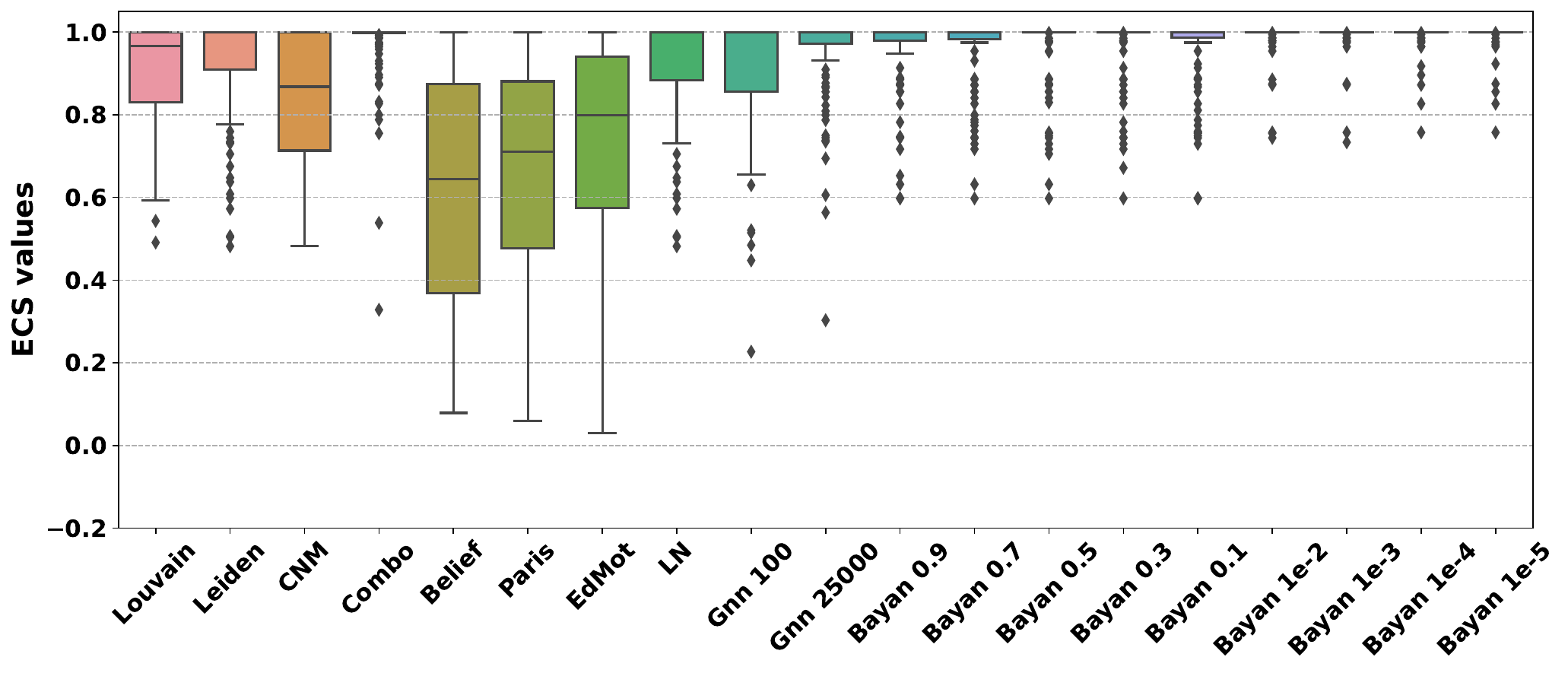}
 \end{subfigure}
	\caption{The box plots of similarity to an optimal partition (AMI in the top panel, RMI in the center panel, and ECS in the bottom panel) for each algorithm based on all 104 network instances. (Magnify the high-resolution figure on screen for more details.)}
 \label{fig:distribution}
\end{figure}

In the scatter-plots Figures~\ref{fig:GOP_AMI}--\ref{fig:GOP_ECS}, data points overlap with each other and therefore distributions are not visible. Figure~\ref{fig:distribution} complements the observations made from Figures~\ref{fig:GOP_AMI}--\ref{fig:GOP_ECS}. Figure~\ref{fig:distribution} illustrates for each algorithm the box plots of AMI, RMI, and ECS values, obtained on all 104 networks. Each box in Figure~\ref{fig:distribution} shows: the first quartile ($Q_1$), the median ($Q_2$), and the third quartile ($Q_3$) of the distribution for one similarity measure and one algorithm. The whiskers are drawn from the nearest hinge ($Q_1$ or $Q_3$) to the farthest datapoint within the 1.5 interquartile range ($\pm 1.5(Q_3-Q_1)$). 

 The distributions for the three measures AMI, RMI, and ECS are quite similar reaffirming that the differences between algorithms observed in Subsection \ref{ss:scatter} are irrespective of the choice of partition similarity measure. The alignment between our AMI and RMI results is consistent with the results in \cite{jerdee2023normalized} while that study recommends using RMI. 
 
 Belief, Edmot, and Paris have the three widest distributions for all three similarity measures. The medians of all three measures are below $0.8$ for Paris indicating that its failure to return partitions similar to optimal happens on half of these instances. The median ECS for Belief is also below $0.8$ which can be interpreted similarly. For CNM and EdMot, the medians are around 0.85 and 0.9 showing the same issue but to a lesser degree. 
 
 All the distributions are left-skewed indicating higher variability among values below the median. Compared to the other heuristics, Louvain, Leiden, LN, and Combo have distributions with smaller ranges and higher medians. Both variations of the GNN algorithm have wider distributions than Combo. The nine variations of the Bayan algorithm have medians extremely close to 1 with some of them also having the smallest ranges among all the algorithms considered. We reobserve the expected pattern that reducing the approximation threshold of Bayan (from $0.9$ to $1e-5$) generally leads to better performance (higher similarity to an optimal partition with lower variation).

\FloatBarrier

\subsection{Empirical time comparison of the algorithms}
\label{ss:time}

\begin{figure}[!htbp]
\centering
	\includegraphics[width=1\textwidth]{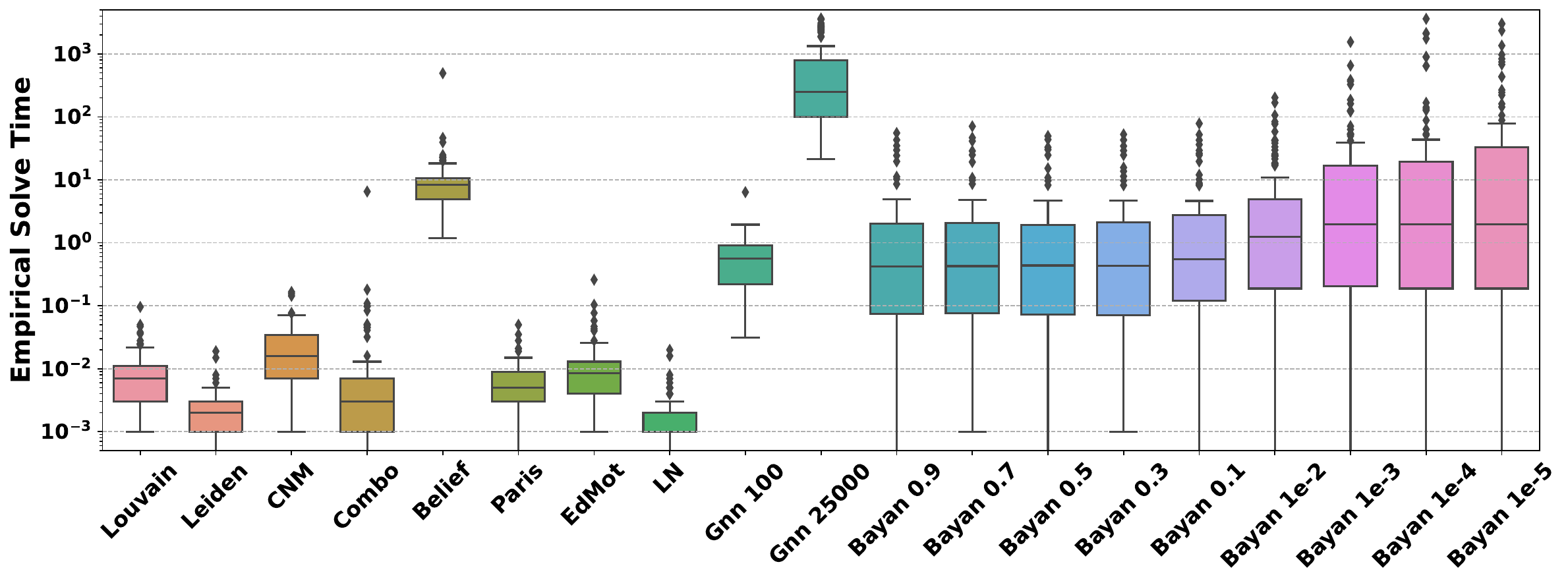}
	\caption{Box plots representing the empirical solve time of the algorithms for the 104 networks. The y-axis is in logarithmic scale. (Magnify the high-resolution color figure on screen for more details.)}
	\label{fig:time}
\end{figure}

All these algorithms attempt to solve the same optimization problem: maximizing modularity, but their ways of exploring the feasible space are widely different leading to considerably different solve times for their specific computations (which are inherently different). Figure~\ref{fig:time} shows for each algorithm a box plot of its empirical solve times measured on the 104 networks. Note that the y-axis of Figure~\ref{fig:time} is on a logarithmic scale. These solve times are empirical and depend on the computing resources used, but are comparable to each other because the same computing resources are used for all these algorithms.

Box plots of two algorithms stand out: GNN-25K has the longest solve time (and a median of 247 seconds) followed by the Belief algorithm (that has a median of 8.4 seconds). The median solve time of GNN-100 (0.56 seconds) and that of five Bayan variations are similar (those with approximation tolerances from $0.1$ to $0.9$). Other variations of Bayan have a median solve time above one second with the slowest variation (Bayan $1e-5$) having a median of 1.97 seconds. The distributions of solve times for all variations of Bayan are left-skewed with their $Q_1$ solve time often being smaller than their median solve time by an order of magnitude. The widest boxes belong to different variations of Bayan for which the $Q_3$ often an order of magnitude larger than $Q_2$. Except for Belief, the heuristic algorithms are 1-2 orders of magnitude faster than GNN and Bayan; with Leiden and LN being the fastest algorithms in our analysis. 

The solve time distributions of all algorithms have outliers (values larger than the top whisker of the box plot). All 104 instances considered are from the networks with tens to a few thousands of edges. Therefore, the outliers existing for almost all algorithms indicates that solving some instances take much longer the typical solve time of that algorithm for that range of input size. This can be partially explained by the differences in graph structures. Some graphs have a structure close to the structure of a random graph (and far from a modular structure). Finding a high-modularity partition for them takes orders-of-magnitude longer (than the typical time for networks of that size range) irrespective of the algorithm used.

\begin{figure}[!htbp]
\centering
	\includegraphics[width=0.8\textwidth]{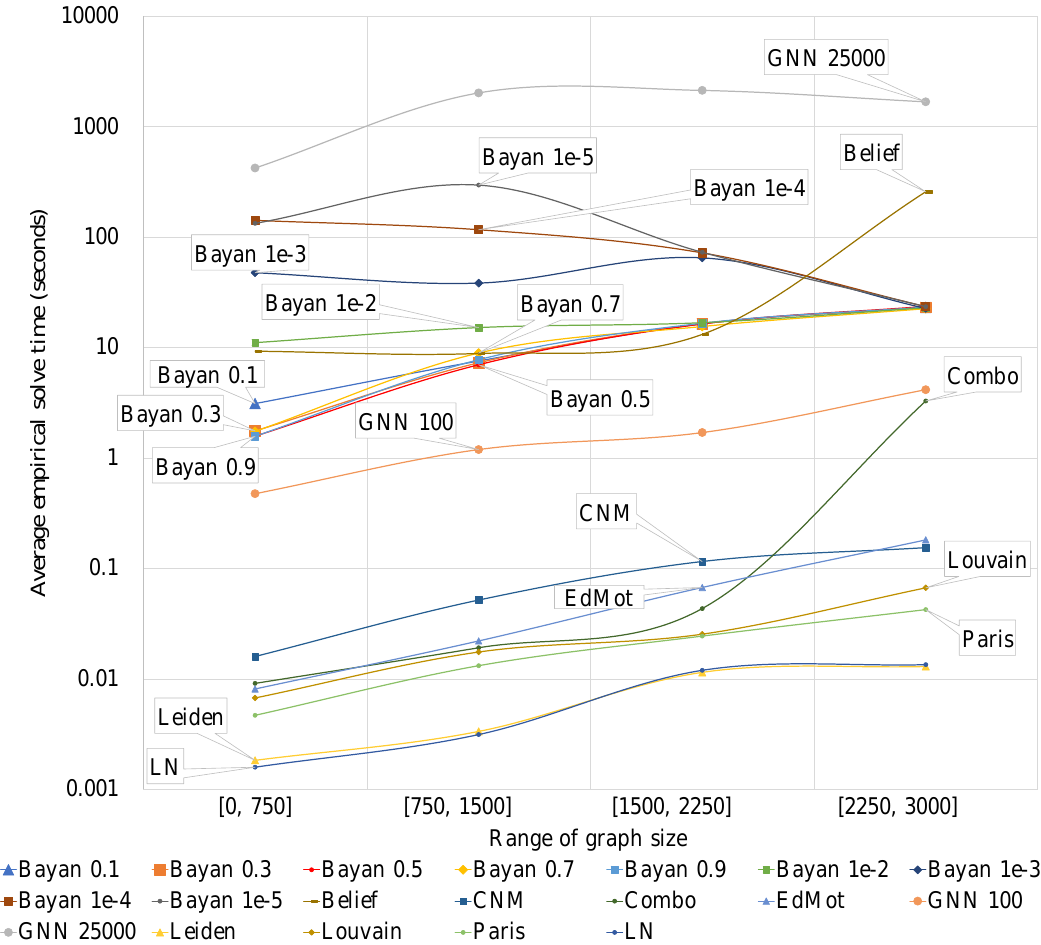}
	\caption{Average solve time of different methods based on bins of input size. The y-axis is in logarithmic scale. (Magnify the high-resolution color figure on screen for more details.)}
	\label{fig:scaling}
\end{figure} 

Figure~\ref{fig:scaling} shows the same solve times, but with different arrangement and visualization. It shows how the average empirical solve time of each algorithm changes as the input size increases. For Figure~\ref{fig:scaling}, we group the 104 networks into four bins based on their graph sizes (number of edges) and plot the average solve time of each algorithm for each of the four bins of instances. Note that the y-axis in Figure~\ref{fig:scaling} is in logarithmic scale. It can be observed that GNN-25k has the highest solve time among all algorithms; on average, it takes over 1000 seconds for graphs with more than 750 edges. 

The solve time of these algorithms are generally increasing with input size, but Figure~\ref{fig:scaling} does not show a monotonic increase for some methods like Bayan 1e-5 and Bayan 1e-4. In case of Bayan, this can be partly explained by considering that some larger instances in our analysis have modular structures that have facilitated the exploration of Bayan for approximating an optimal solution. Different variations of Bayan, Belief, and GNN-100 take order-of-magnitude longer than other modularity-based algorithms. At the other extreme, LN and Leiden are the fastest among all algorithms in our analysis. Counting the horizontal gridlines in Figure~\ref{fig:scaling}, there are five orders of magnitude difference between the slowest (GNN-25k) and the fastest (Leiden and LN) algorithms for processing instances of comparable sizes.

\FloatBarrier

\subsection{Success rates of the algorithms in maximizing modularity}\label{ss:success}

The GOP values help us answer a fundamental question about these modularity-based algorithms: how often does each algorithm return an optimal partition? We report the fraction of networks (out of 104) for which a given algorithm returns an optimal partition. A similar assessment of some of these algorithms on a different set of networks is provided in \cite{aref2023suboptimality}.

\begin{figure}[!htbp]
    \centering
    \includegraphics[width=0.65\textwidth]{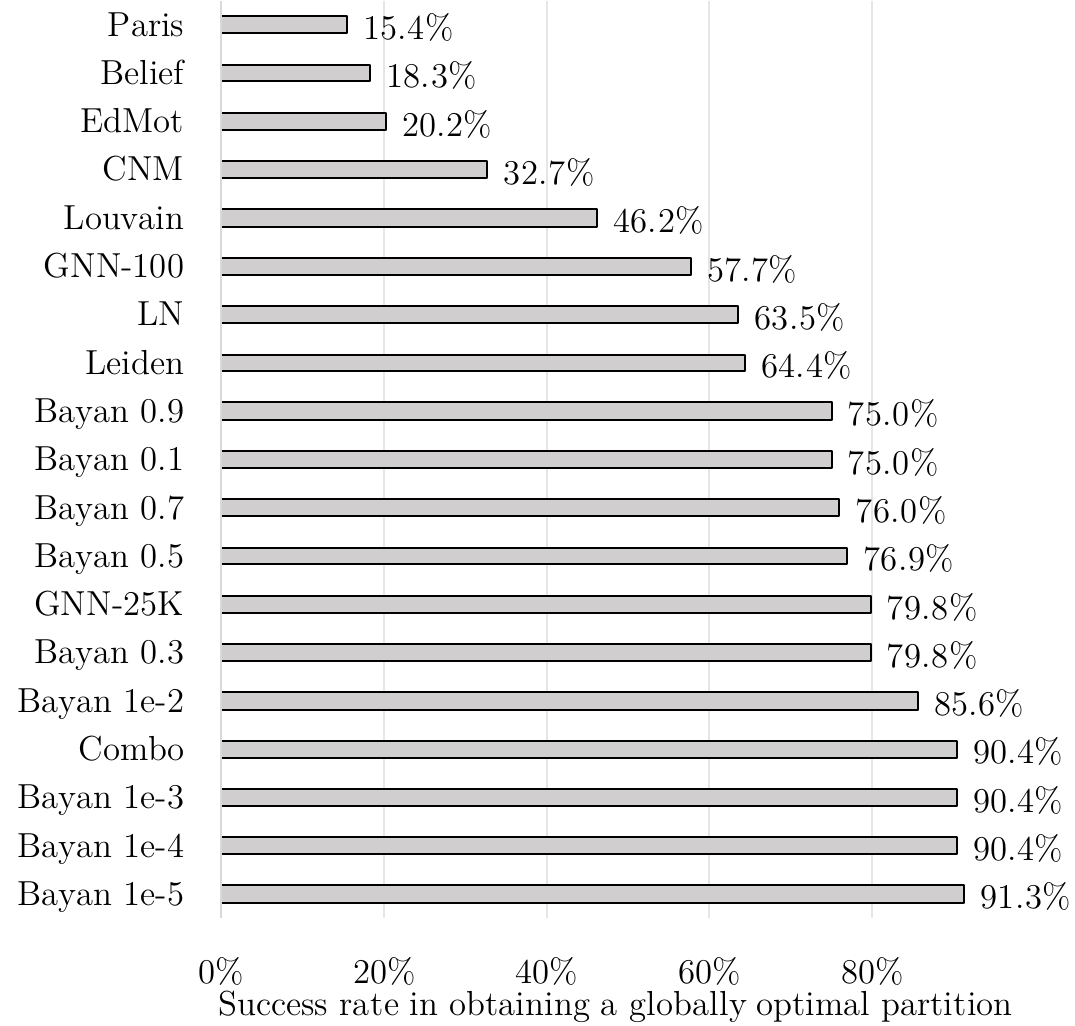}
    \caption{Ten algorithms and their variations ranked based on their success rate in achieving global optimality on the 104 networks considered.}
    \label{fig:success}
\end{figure}

Figure~\ref{fig:success} shows the success rates of all algorithms and their variations in achieving global optimality on the 104 networks considered. Among the eight heuristics, Combo has the highest success rate, returning an optimal partition for 90.4\% of the networks. The average success rate of all 8 heuristics combined is 43.9\%. With the exception of Combo, the rates for these heuristics are arguably low success rates for what the name \textit{modularity maximization algorithm} implies or the idea of discovering network communities through maximizing a function. The two variations of GNN have markedly different success rates and their average success rate is 68.7\%. The least restricted version of approximate Bayan (Bayan 0.9) returns optimal partitions on 75\% of networks and is more successful than the average heuristic and the average GNN variation. Bayan's success rate increases to 91.3\% when a sufficiently small approximation threshold is chosen. This indicates that in the context of solving this NP-hard graph optimization problem, an optimization procedure (branch and bound which is used in Bayan) can be more successful than other existing alternatives (e.g.,\ heuristics and GNNs). After all, modularity maximization is a mathematical optimization task and it should be safe to assume that guaranteed approximate optimization methods would likely be hard to outperform in optimality success rate (acknowledging that they most likely take longer and can only be used for small and mid-sized networks). Figure~\ref{fig:success} also shows that using Bayan with a smaller approximation tolerance value often leads to a higher success rate. 

Earlier in Figures~\ref{fig:GOP_AMI}--\ref{fig:GOP_ECS}, we observed that near-optimal partitions tend to be disproportionately dissimilar to any optimal partition. In other words, close-to-maximum modularity partitions are rarely close to any optimal partition. Taken together with the low success rates of the average heuristic in maximizing modularity, our results indicate a crucial mismatch between the design philosophy of these methods and their average capability: most heuristic modularity maximization algorithms rarely return an optimal partition or a partition resembling an optimal partition even on graphs with modular structures.

\section{Discussions and Future Directions}

Understanding modularity capabilities and limitations has been complicated by the under-explored performance of modularity maximization algorithms on the task that is literally in their name: modularity maximization. Previous methodological studies \cite{lancichinetti_limits_2011,Chen_2018,global2018,peixoto_2023}, which have shed light on other aspects, have rarely disentangled the unguaranteed aspect of the inexact optimization from the fundamental concept of modularity. Our study is a continuation of previous efforts \cite{good_performance_2010} in separating the effects of sub-optimality (or the choice of using greedy algorithms \cite{kawamoto2019counting}) from the effects of using modularity on the fundamental task of detecting communities. 

We analyzed the effectiveness of eight heuristics \cite{clauset_finding_2004,blondel_fast_2008,rb_pots_2008,sobolevsky2014general,zhang2014,paris_2018,traag_louvain_2019,edmot_2019}, two variations of a GNN \cite{sobolevsky2022gnn}, and several variations of an approximation algorithm \cite{aref2022bayan} in maximizing modularity. While our findings are limited to only a handful of algorithms, their combined usage by tens of thousands of peer-reviewed studies \cite{Kosowski2020} motivates the importance of this assessment. Most heuristic algorithms for modularity maximization tend to scale well for large networks \cite{zhao2021community}. They are widely used not only because of their scalability or ease of implementation \cite{kawamoto2019counting}, but also because their high risk of algorithmic failure is not well understood \cite{kawamoto2019counting}. The scalability of these heuristics comes at a cost: their partitions have no guarantee of proximity to an optimal partition \cite{good_performance_2010} and, as our results showed, the average heuristic rarely returns an optimal partition even on modular graph structures. Moreover, we showed that their sub-optimal partitions tend to be disproportionately dissimilar to any optimal partition irrespective of the partition similarity metric chosen. 

Our approach of quantifying similarity between partitions using AMI, RMI, and ECS has some merits over our earlier study \cite{aref2023suboptimality}, but it still has some limitations. An alternative approach involves identifying \textit{building blocks} across different candidate partitions of a network \cite{Riolo_building_block}. These building blocks are groups of nodes that are usually found together in the same community. Riolo and Newman propose a method for finding building blocks and suggest that building blocks obtained in their results are largely invariant for a given network while different arrangements of the same building blocks lead to different partitions of that network \cite{Riolo_building_block}. While this is a viable explanation of some of the variations between some community detection algorithms, our results do not fully match with this interpretation of partition dissimilarities. For example, the communities (shown by colors) in Fig. \ref{fig:facebook1} can only be interpreted as a specific arrangement of very small building blocks including blocks made up of a single node. In this case with such building blocks sizes, finding a suitable arrangement of these extremely small building blocks is arguably the whole task of community detection on which modularity-based algorithms perform differently.

Neither using modularity nor succeeding in maximizing it is required for CD at the big-picture level. A common narrative in the literature is debating whether modularity is suitable or not \cite{Miasnikof2020,peixoto_2023}. We argue that such a debate is an oversimplification because suitability of using modularity depends on the task\footnote{see \cite{schaub2017many} for four different tasks that are all referred to as community detection.}, the context, and several other factors including (1) whether it used as an objective function \cite{dinh_toward_2015} or as a partition quality function \cite{Miasnikof2020}; (2) how the maximization is operationalized; and (3) what advantages are offered by the alternatives to use instead of modularity. Our results shed light on the first two questions, but are not related to alternative methods that do not use modularity. A recent study claims modularity maximization is the most problematic CD method and considers it harmful \cite{peixoto_2023}. Another study shows that, given computational feasibility, exact maximization of modularity outperforms 30 other CD methods in accurate and stable retrieval of ground-truth communities in both LFR and ABCD benchmarks \cite{aref2022bayan} suggesting the relevance of modularity for CD. 

Our results were based on small and mid-sized networks with no more than 2812 edges. We showed the extent to which each modularity-based method succeeds in returning optimal partitions or partitions resembling optimal partitions. Given that modularity maximization is an NP-hard problem, it is not reasonable to expect that the performance of these inexact methods suddenly increases for large-scale networks. The extent of their failures on large-scale networks is not quantified yet. However, the average success rate of 43.9\% suggests the performance in maximizing modularity that can be expected from these heuristics on other networks. This expectation is realistic for small and mid-sized networks, and it is arguably optimistic for large-scale networks.

Our findings suggest that if modularity is to be used for detecting communities, using approximation \cite{cafieri2014reformulation,dinh_network_2015,kawase2021,aref2022bayan} and exact \cite{aloise_column_2010,aref2022bayan,brusco_maximization_2023} algorithms is recommendable for a more methodologically sound usage of modularity within its applicability limits. 

A promising path forward could be using the advances in integer programming to develop better approximation algorithms (outperforming approximate Bayan) for solving the mathematical models of modularity maximization \cite{brandes2007modularity,agarwal_modularity-maximizing_2008,dinh_toward_2015} on networks of practical relevance within the limits of computational feasibility.
New heuristic algorithms that strike a balance between accurate computations and scalability (achieving or surpassing the performance of methods like Combo and Leiden, but with higher scalability) may also be useful particularly for large-scale networks.

\subsubsection{Author contributions}
Conceptualization (SA); data curation (SA); formal analysis (SA, MM); funding acquisition (SA); investigation (SA); methodology (SA, MM); project administration (SA); resources (SA, MM); software (SA, MM); supervision (SA, MM); validation (SA, MM); visualization (SA, MM); writing - original draft preparation (SA); writing - review \& editing (SA, MM).

\subsubsection{Acknowledgements} Authors are thankful to the three anonymous referees of ICCS-2023 and to the two anonymous referees of this special issue for their helpful comments on earlier version of this article. We also thank Max Jerdee for providing his RMI code and we acknowledge Santo Fortunato, Mark Newman, and Tiago P. Peixoto for the helpful correspondence and discussions. 

\section*{Appendix}

\subsection*{Justifying the use of AMI, RMI, and ECS and the exclusion of other popular measures of partition similarity}

We use AMI, RMI, and ECS because they are shown to be reliable measures of partition similarity \cite{vinh_AMI,gates2019element,newman2020RMI,jerdee2023normalized}.

We avoid using NMI because, despite its common use, several studies indicate that using NMI leads to incorrect assessments \cite{vinh_AMI,newman2020RMI,gates2019element} and incorrect evaluation of competing algorithms \cite{jerdee2023normalized}. For example, consider two arbitrarily dissimilar partitions A and B, each made up of two communities and their NMI taking the value 0.1 (for the sake of argument). If we take partition B and split each of its communities into several communities, the NMI between the resulting partition and partition A will increase; and this increase is monotonic with the more we split partition B \cite{gates2019element}. Despite that this undesirable bias of the NMI towards the number of clusters has been well documented for several years \cite{gates2019element}, NMI is still among the most popularly used \cite{roozbahani_community_2023,singh_disintegrating_2022,hamid_fast_2018,khomami_new_2018,sattari_cascade_2018,schumm_bloom_2012} metrics for quantifying the similarity of partitions \cite{jerdee2023normalized}.

Similar to the NMI, other popularly used measures of partition similarity suffer from at least one form of undesirable bias \cite{gates2019element}.  
We also avoid using any of the four measures: the Jaccard index, the Fowlkes-Mallows index, the adjusted Rand index, and the F measure because they suffer from a bias that favors skewed cluster sizes \cite{gates2019element}. 

\subsection*{Accessing the data for real and synthetic networks}

The data on the 50 LFR and ABCD graphs used in this study are available in a \textit{FigShare} data repository \cite{Aref2023figshare_modular}. 

The LFR benchmarks used in this study were randomly generated based on the following parameters: number of nodes ($n$) randomly chosen from the range $[20,300]$, maximum degree $\lfloor 0.3n \rfloor$, maximum community size $\lfloor 0.5n \rfloor$, power law exponent for the degree distribution $\tau_1=3$, power law exponent for the community size distribution $\tau_2=1.5$, and average degree of 4. The parameter $\mu$ (LFR mixing parameter) was chosen from the set $\{0.01, 0.1\}$ (10 LFR graphs for each value of $\mu$).

The ABCD benchmarks were randomly generated based on the following parameters: number of nodes ($n$) randomly chosen from the range $[10, 500)$; minimum degree $d_{min}$ and minimum community size $k_{min}$ randomly chosen from the range $[1, n/4)$; maximum community size chosen randomly from $[k_{min} + 1, n)$; maximum degree chosen randomly from $[d_{min} + 1, n)$; and power law exponents for the degree distribution and community size distribution randomly from $(1, 8)$ and then rounded off to 2 decimal places. The parameter $\xi$ (ABCD mixing parameter) was chosen from the set $\{0.01, 0.1, 0.3\}$ (10 ABCD graphs for each value of $\xi$). 

The 54 real networks were loaded as simple unweighted and undirected graphs. They are available in the publicly accessible network repository \href{https://networks.skewed.de/}{Netzschleuder} with the 54 names below:

dom, packet\_delays, sa\_companies, ambassador, florentine\_families, rhesus\_monkey, kangaroo, internet\_top\_pop, high\_tech\_company, moviegalaxies, {november17}, moreno\_taro, sp\_baboons, bison, dutch\_school, zebras, cattle, moreno\_sheep, 7th\_graders, college\_freshmen, hens, freshmen, karate, dutch\_criticism, montreal, ceo\_club, windsurfers, elite, macaque\_neural, sp\_kenyan\_households, contiguous\_usa, cs\_department, dolphins, terrorists\_911, train\_terrorists, highschool, law\_firm, baseball, blumenau\_drug, lesmis, sp\_office, polbooks, game\_thrones, football, football\_tsevans, sp\_high\_school\_new, revolution, student\_cooperation, interactome\_pdz, physician\_trust, malaria\_genes, marvel\_partnerships, facebook\_friends, netscience

For more information on each network and its original source, one may check the Netzschleuder website by adding the network name at the end of the url:  https://networks.skewed.de/net/. For example, \url{https://networks.skewed.de/net/malaria_genes} provides additional information for the \textit{malaria\_genes} network. In cases of multiple networks existing with the same name in Netzschleuder, we have only used the lexicographically first network (e.g. we have only used the \textit{HVR\_1} network from \url{https://networks.skewed.de/net/malaria_genes}).

\end{document}